\documentclass[12pt,aos]{article}
\usepackage{epsf}
\usepackage{amsthm,amsmath,amssymb,graphicx,graphics,epsfig}
\usepackage{booktabs}
\usepackage{multirow}
\usepackage{lineno}
\usepackage{hyperref}
\usepackage[authoryear,round]{natbib}
\usepackage{algorithm}
\usepackage{algorithmic}
\usepackage{graphicx}
\usepackage{float}
\usepackage{subfigure}
\usepackage{natbib}
\usepackage{color}
\usepackage[authoryear,round]{natbib}
\setcitestyle{numbers,square}

\newtheorem{theorem}{Theorem}[section]

\newcommand{\bi}{\begin{itemize}}
\newcommand{\ei}{\end{itemize}}
\def\var{\mbox{var}}

\def\E{\mbox{E}}

\def\f{(X_i,Y_i)_{i=1}^m}
\def\mbeta{\hat\beta_g}
\def\p{\mbox{pr}}

\def\diag{\mbox{diag}}

\def\tbeta{\tilde{\beta}_r}

\begin{document}
\baselineskip 8mm
\setcounter{page}{0}
\bibliographystyle{plainnat}

\begin{center}
{\Large \bf Optimal subsampling algorithm for the marginal model with large longitudinal data}
\end{center}
\vspace*{2mm}

\begin{center}

{Haohun Han$^{a}$, ~Liya Fu$^{a,*}$}\vspace*{0.3cm}\\

\normalsize{ \it $^a$School of Mathematics and Statistics, Xi'an Jiaotong University, China}\\
{\it $^*$Email: fuliya@mail.xjtu.edu.cn}

\end{center}

\centerline{\bf ABSTRACT}

Big data is ubiquitous in practices, and it has also led to heavy computation burden. 
To reduce the calculation cost and ensure the effectiveness of parameter estimators, an optimal subset sampling method 
is proposed to estimate the parameters in marginal models with massive longitudinal data. 
The optimal subsampling probabilities are derived, and the corresponding asymptotic properties are established to ensure the 
consistency and asymptotic normality of the estimator. Extensive simulation studies are carried out to evaluate the
performance of the proposed method for continuous, binary and count data and with four different working correlation matrices.
A depression data is used to illustrate the proposed method.


\section{Introduction}  
   Big data is becoming more and more common in practices. How to reduce the computation burden
   is a very important topic, especially when the amount of data 
   is too large for computers to handle. 
   An effective approach to solve this problem is the optimal subsampling methods \citep{ai2021,drin11,wang19a,wang18b}. 
   \citet{drin06}  presented a sampling algorithm for the linear-algebraic problem of the least-squares fit. 
   For linear regression models with the massive data, \citet{ma15} proposed a robust and efficient optimal subset method  based on the
    regularised leverage score.
    \citet{wang18a} constructed an information-based optimal subdata selection (IBOSS) method, which accelerates the calculation speed and is suitable for the distributed parallel computing. 
    \citet{wang19a}  developed a divide and conquer IBOSS method to solve the situation where the number of the volumes in a dataset is very large.
     For the logistic regression models, \citet{wang18b}  established the
      asymptotic properties of the estimators and then proposed an A-optimization subsampling strategy by minimizing the trace of the asymptotic covariance matrix. \citet{zuo2021} proposed a distributed subsampling procedure and constructed effective subsample-based estimators for logistic models with the massive and distributed data.
   \citet{ai2021} studied the generalized linear models  with the massive data and constructed the A- and L-optimal criteria to derive the subsampling probabilities.
  \citet{wang:ma:2021} and \citet{fan2021} investigated the   optimal subset sampling methods for quantile regression and provided the optimal subsampling probabilities, respectively. Furthermore, \citet{yu2022} studied the mean regression and derived optimal Poisson subsampling probabilities
for maximum quasi-likelihood estimation under the A- and L-optimality criteria.
   \citet{yuan2022} studied the  composite  quantile regression with big data and  obtained the optimal subsampling probabilities under the A- and L-optimality criteria.
   All the methods mentioned above focus on the independent data.

   Longitudinal data are very common in  medical research and economics studies.   To our knowledge, there is little work on the optimal subsampling for the massive longitudinal data. 
 \citet{wang2023} firstly studied the optimal subsampling for the linear regression with the massive longitudinal data. They   combined the leverage-based and gradient-based subsampling methods and developed a new optimal sampling method under the assumption that the random errors follow a multivariate normal distribution with an exchangeable correlation matrix. 
 
The marginal models mainly make statistical inference on the population-average  in longitudinal data analysis \citep{diggle2002},
which is suitable for the discrete data and continuous data. Furthermore, there is 
 no  assumption on the specific joint distribution of the response variables.
 \citet{lian86} proposed a milestone  generalized estimating equations (GEE) method for parameter estimation in the marginal model, which incorporates the within correlations, and the  parameter 
   estimators  are consistent even the  correlation matrix is misspecified.
 For a given number of observations for each subject, \citet{crow86} proved the weak convergence of the GEE estimator.
 \citet{li96} gave the asymptotic normality conditions and the rate of convergence of the estimator based on the minimax method proposed by \citet{cram46},
  and \citet{yuan98} gave the existence, consistency, and asymptotic normality of the estimator under some general conditions.

  However, the computational cost 
  of GEE increases sharply with the increase of the sample size and the number of observations for each subject.
Therefore, in this paper, we propose  an optimal subset sampling method for the marginal model with large longitudinal data. We provide weighted generalized estimating equations and prove the consistency and asymptotic normality of the resulting estimator. The asymptotic properties do not depend on the specific working correlation structure. Moreover, we derive the optimal subsampling probabilities  by minimizing the trace of the covariance matrix of the regression estimators. 

   The rest of the paper is organized as below. 
   In Section 2, the marginal model and some notation are introduced. In Section 3, the asymptotic properties are established, and the optimal subsampling
   probabilities are given. In Section 4, numerical simulations are carried out to evaluate the performance of the proposed method.
  In Section 5, a practical data set is analyzed. 
  In Section 6, some conclusions are summarised, and the future work is discussed.
   The proofs of the theorems are presented in the supplemental material.

\section{The marginal model}

Suppose $(y_{ij},x_{ij})$ is the $j$th measurement of the $i$th subject for
 $j=1,2,\cdots, n_i$ and $i=1,2,\cdots,m$, where
$y_{ij}$ is the response, and $x_{ij}$ is a $p\times 1$ corresponding covariate vector.
Without loss of generality, we assume that each individual has the same observations, that is $n_1=\cdots=n_m=n$. The expectation and the variance function of $y_{ij}$ are given by
$$
\mu_{ij}=\E(y_{ij} | x_{ij})=g(\eta_{ij})\quad \mbox{and} \quad \var(y_{ij} | x_{ij})=\nu_{ij}=\phi\nu (\mu_{ij}),
$$
 where $g(\cdot)$ is a known  differentiable link function and $\eta_{ij}=x_{ij}^T\beta$, in which $\beta \in \mathbb{R}^p$ is an unknown parameter vector. 
 When $g(\cdot)$ is the identify link function and logistic function, the model is corresponding to a linear regression 
 and a logistic regression for the longitudinal data, respectively.

Let $Y_i=(y_{i1},\cdots,y_{in})^T$, $X_i=(x_{i1},\cdots,x_{in})^T$ and mean vector $\mu_i=(\mu_{i1},\mu_{i2},\cdots,\mu_{in})^T$. 
Assume that
the working correlation matrix of $Y_i$ is $R(\rho)$ with a $q$-dimensional unknown parameter vector $\rho$. The corresponding working covariance matrix of $Y_i$ is $W_i=\phi A_i^{1/2}R(\rho)A_i^{1/2}$,  where 
$\phi$ is dispersion parameter, and $A_i=\diag(\nu(\mu_{i1}),\cdots,\nu(\mu_{in_i}))$ is a diagonal matrix.
The generalized estimating equation proposed by \citet{lian86} are expressed as
\begin{eqnarray}\label{gee}
 U(\beta)=\sum_{i=1}^m D_i^T A_i^{-1/2}R^{-1}(\rho)A_i^{-1/2} S_i=0,
 \end{eqnarray}
where
$D_i=\partial \mu_i/\partial \beta$,
 and $S_i=Y_i-\mu_i(\beta)$.

 Let $\mbeta$ be the estimator obtained from equation  (\ref{gee}). 
 Generally, there is no closed-form  of $\mbeta$,
 and a modified Fisher scoring algorithm  is usually used to obtained the estimator numerically.
Given a current estimate $\hat\rho$ and $\hat\phi$ of the nuisance parameters $\rho$ and $\phi$, the following 
 iterative procedure can be used to  obtain $\mbeta$:
 \begin{small}
\begin{eqnarray}\label{gee1}
\hat\beta^{(k+1)}_g=\hat\beta^{(k)}_g-\left\{\sum_{i=1}^m D_i^{\rm T}(\hat\beta^{(k)}_g)W_i^{-1}(\hat\beta^{(k)}_g)D_i(\hat\beta^{(k)}_g)\right\}^{-1}\left\{\sum_{i=1}^m D_i^{\rm T}(\hat\beta^{(k)}_g)W_i^{-1}(\hat\beta^{(k)}_g)S_i(\hat\beta^{(k)}_g)\right\},
\end{eqnarray}
\end{small}
where $W_i(\beta)=W_i(\beta,\hat\rho(\beta,\hat\phi))$. The complete iterative algorithm is shown in Algorithm \ref{alg1}.
According to \cite{lian86},  $\mbeta$ is consistent even if the correlation matrix $R(\rho)$ is misspecified.
When $R(\rho)$ is the identify matrix, we obtain the independence estimating equations.
\begin{algorithm}
	\caption{The iterative  algorithm for obtaining $\mbeta$ via equation (\ref{gee}).} \label{alg1}
 \begin{algorithmic}
\STATE {\bf Step 1.} Given an initial estimate of $\beta$, such as $\hat\beta_I$ obtained from the independence estimating equations.

\STATE {\bf Step 2.} Obtain an estimate $\hat\rho$ of the  correlation parameter $\rho$ in the working correlation matrix $R(\rho)$ based on the current estimate $\mbeta^{(k)}$.

\STATE {\bf Step 3.} Update the estimate  $\mbeta^{(k+1)}$ via equation(\ref{gee1}).

\STATE {\bf Step 4.} Repeat Steps 2--3 until the estimate $\mbeta^{(k+1)}$ satisfies certain convergence criterion.

 \end{algorithmic}
\end{algorithm}

In each iteration of equation(\ref{gee1}),  it requires $O(Mp^2)$ computing time, where $M$ is the total number of observations.
To obtain a consistent estimate of $\beta$, it need calculate  at least $O( c\cdot Mp^2)$ times, where $c$ is the number of iteration.
When the sample size $m$ is large, the computing burden is heavy.
Therefore, computation is a bottleneck for the application of the marginal model on the massive longitudinal data. 
To reduce the computational cost, an optimal subsampling method will be considered in the following section. 

\section{The optimal subsampling}
In this section, we first present a general subsampling algorithm, and then establish
the asymptotic properties of the resulting estimator, and finally provide an optimal subsampling 
strategy to reduce the computational burden.

\subsection{Subsample-based estimator and its asymptotic properties}

Now, sample with replacement from the original $m$ data points $r$ times with probabilities
$\pi=\{\pi_i\}_{i=1}^m$ to form the subsample set $\mathcal {D}=(X_i^{*},Y_i^{*},\pi_i^*)_{i=1}^r$, where $\pi^*_i$ corresponds to the sampling probability of  the subset  of $(X_i^{*},Y_i^{*})$.
Based on the subsample set $\mathcal {D}$, calculate the weighted GEE:
\begin{eqnarray}\label{sgee}
 U^*(\beta)=\frac{1}{r}\mathop\sum_{i=1}^r \frac{1}{\pi_i^*}D_i^{*T} W_i^{*-1} S^*_i=0,
\end{eqnarray}
where $D_i^*=\partial{\mu_i^*}/\partial{\beta}$ with $\mu_i^*=g(X_i^*\beta)$, and $W_i^*$ is the working covariance matrix of $Y_i^*$, and $S^*_i=Y^*_i-\mu^*_i(\beta)$.
We can solve the above equation (\ref{sgee}) and obtain an estimator $\tbeta$ based on the subsampling observations.

To prove asymptotic properties of the subsampled
estimator $\tbeta$, we require the following  regularity conditions.

(i) $\eta$ lies in the interior of a compact set $K\in\mathbb{R}^n$ almost surely,
where $\eta$ is the random variable with the same distribution as $\eta_i$.

(ii) Let
$\dddot{\mu}(\eta)$ be  the third
derivatives of $\mu(\eta)$, and assume $\dddot{\mu}(\eta)$ is continuous with respect to $\eta$ in $K$.

(iii) 
$\nu(\mu)\ge C$ for some $C>0$ and any $\mu \in \mu(K)$.

(iv) As $m\to \infty $, for $\gamma=0$ and some $\gamma >0$,
$m^{-(2+\gamma)}\mathop\sum_{i=1}^m  \|X_i\|_2^{2+\gamma} \| S_i\|_2^{2+\gamma}/\pi_i^{1+\gamma}=O_p(1)$.

(v) As $m\to \infty $, $m^{-1}\mathop\sum_{i=1}^m  \|X_i\|_2^{2} \| S_i\|_2^{2}=O_p(1)$ and $m^{-1} \sum_{i= 1}^{m}\left\|Y_{i}-\mu_i\right\|_{1}^{2}=O_p(1)$.

(vi) As $m\to \infty $, $\E\|X_i\|_2^6=O_p(1)$ and $m^{-2}\mathop\sum_{i=1}^m\|X_i\|_2^4/\pi_i=O_p(1)$.

(vii) Let $N=mn^{-1/2}$, as $m\to \infty $, $\Phi_0=N^{-1}\partial U(\beta)/ \partial \beta\left.\right|_{\beta=\mbeta} $ goes to a positive-definite matrix in probability. 

(viii) $\inf_{\beta:\|\beta-\mbeta \| \ge \epsilon }\| U(\beta)\|>0 $ for any $\epsilon >0$.

(ix) Let $R^{-1}=(w_{ij})_{n\times n}$, and assume there exists some $C_{\rho}>0$, $\mbox{max} |w_{ij}|<C_{\rho}$
 for any $\rho \in \Theta_{\rho}$, where $\Theta_{\rho}$ is the nature parameter space of $\rho$.

 Here, conditions (i), (ii), (vii) and (viii) guarantee the estimator obtained by the GEE is consistent
 for the full data. Conditions (iii) and (iv) are necessary for the Lindeberg-Feller
 central limit theorem, and they guarantee the first and second moments of the estimator obtained 
 by subsample is existed.
 Conditions (v) and (vi) can be interpreted as the requirement for the distribution of the error term,
 requiring that the variance of the error term exists. The last condition (ix)
 is to exclude the case that multiple observations from the same subject are the same. 
 This is because Theorem \ref{th2} requires the order of the subsample size, 
 and if multiple observations from the same subject were the same, it would make the subsample size  much smaller than that
  we expect, which would not guarantee the asymptotic normality of the proposed estimator.

\begin{theorem}\label{th1}
  Under the assumptions (i)--(ix), for any $\epsilon>0$, there exists some $C>0$ and $r_0>0$, such that
 $$
    \p(\| \tbeta -\mbeta\|\ge r^{-1/2}C|\f)<\epsilon.
$$

\end{theorem}

Theorem \ref{th1} presents the consistency of the estimator $\tilde \beta_r$ based on the subsampling data to the estimator $\hat\beta_g$ based on the full data.
The rate of the converge is $r^{-1/2}$.

\begin{theorem}\label{th2}
  Given $(X_i,Y_i)_{i=1}^m$, under the assumptions (i)-(ix), as $r\to \infty$, $n^2/r \to 0$ and $m\to \infty$,
  \begin{eqnarray*}
     V^{-1/2}(\tbeta-\mbeta)\to \mathcal{N}(0, I),
 \end{eqnarray*}
  where $V=\Phi_0^{-1} \Phi \Phi_0^{-1}$ and
   \begin{eqnarray*}
    \Phi=\frac{1}{r{ N^2}}\mathop\sum_{i=1}^m \frac{D_i^{\rm T}(\mbeta)W_i^{-1}(\mbeta)S_i(\mbeta)S_i^{\rm T}(\mbeta)W_i^{-1}(\mbeta)D_i(\mbeta)}{\pi_i}.
  \end{eqnarray*}
\end{theorem}
Theorem \ref{th2} indicates that the error of  $\tilde \beta_r$ and  $\hat\beta_g$  is asymptotically normal, and
the asymptotic distribution of $\tilde \beta_r$ depends on the subsampling probabilities $\{\pi_i\}_{i=1}^m$.
As the sample size or the number of observations from each subject increases, 
the parameter estimators become better, even if the correlation matrix is misspecified. 
Assuming that the dimension of the regression parameters $p$
 is fixed, and that the number of observations from the same subject is large, the number of the correlation parameters in the 
 correlation matrix is $O(n^2)$. Due to the limitations of the GEE model, the sample size is required 
 to be larger than the  parameter dimensionality, which is the another explanation on the relationship between 
 the number of subsamples and the number of observations from the same subject.
 
\subsection{The optimal subsampling probability}

To obtain $\tbeta$ using equation (\ref{sgee}), we need to specify the subsampling probability $\{\pi_i\}_{i=1}^m$
for the full data. We propose minimizing the trace of the covariance matrix $V$ to obtain an optimal subsampling 
probability and denote it  as a mVc-optimal sampling probability, which   means minimizing the summation of the asymptotic variance of each element in $\tbeta$.

\begin{theorem}\label{th3}
The subsampling strategy is mVc-optimal if the subsampling probabilities is chosen such that
\begin{eqnarray}\label{eq:mvc}
\pi_i^{\text{mVc}}=\frac{\left\|\Phi_0^{-1}D_i(\mbeta)^TW_i^{-1}(\mbeta)S_i(\mbeta)\right\|_2}{\sum_{i=1}^m\left\|\Phi_0^{-1}D_i(\mbeta)^TW_i^{-1}(\mbeta)S_i(\mbeta)\right\|_2},\quad i=1,\cdots,m.
  \end{eqnarray}
\end{theorem}
To improve the computational speed, we abandon the $p\times p$ dimension matrix $\Phi_0$ and propose a mV optimal subsampling 
method. 

\begin{theorem}\label{th4}
  The subsampling strategy is mV-optimal if the subsampling probabilities are chosen such that
\begin{eqnarray}\label{eq:mv}
\pi_i^{\text{mV}}=\frac{\left\|D_i(\mbeta)^T W_i^{-1}(\mbeta)S_i(\mbeta)\right\|_2}{\sum_{i=1}^m\left\|D_i(\mbeta)^T W_i^{-1}(\mbeta)S_i(\mbeta)\right\|_2},\quad i=1,\cdots,m.
\end{eqnarray}
\end{theorem}

\noindent\textbf{Remark 1}. According to Theorems \ref{th3} and \ref{th4}, it is easy to find that the denominators in formulas (\ref{eq:mvc}) and (\ref{eq:mv}) can be very close to zero, which makes the error very large and derives singular values in calculating 
the sampling probabilities.
Therefore, in the following algorithm, we impose an lower bound on  $S_i(\beta)$ to circumvent the problem of the denominator being too small.

\noindent\textbf{Remark 2}. In fact, the mV-method does save computation time, but in simulation studies we find that the mean-squared error (MSE) of the estimation obtained in 
this way may be large, especially for the nonlinear models.
Therefore, in the case of non-linear models, the sampling probabilities need to be taken very carefully.

\noindent\textbf{Remark 3}. To solve equation (\ref{sgee}), we need to specify the correlation matrix $R(\rho)$, we consider the 
identify matrix, exchangeable matrix, AR(1) matrix and unstructured matrix in the numerical studies. 
The parameter $\rho$ is estimated by the moment method (\citet{lian86}). 

The optimal sampling probability depends on the estimator $\hat\beta_g$ based on the full data.
Based on Theorem \ref{th3}, we obtain $\tilde{\beta_r}$ via Algorithm \ref{algo2} given as follows:

\begin{algorithm}
  \caption{Optimal subsampling strategy algorithm }\label{algo2}
  \begin{algorithmic}[H]
   \STATE {\bf Step 1.} Given the size of the random subsample $r_0$ and
    set the correlation matrix $R_i= I_{n}$,
     obtain the pilot estimate $\hat\beta_0$ via the GEE.
  \STATE {\bf Step 2.}  Using $\hat\beta_0$, calculate the estimates $\hat\rho$ of the correlation parameters 
  in the working correlation
  matrix $R_i(\rho)$, variance matrix $\tilde{W}_i$ and the optimal probabilities
  $\{\pi_i\}$. To avoid the denominator being too small, we  use $\tilde{S}_{ij}=\max\{S_{ij},\delta\}$ instead 
  of $S_{ij}$, where $\delta $ is a constant number greater than $0$.

  \STATE {\bf Step 3} Sample with replacement for $r$ times based on $\{\pi_i\}$ obtained in Step 2 
  to form the subsampling set. Solve the equation (\ref{sgee}) to
  get $\tbeta$.
  \end{algorithmic}
\end{algorithm}

\section{Simulation studies}
In this section, we  carry out simulation studies to evaluate the performance of the proposed method.
We mainly compare the performance of the sampling methods based on  the following three
sampling probabilities:
$\pi^{\text{Unif}},~~\pi_i^{\text{mVc}}~~ \text{and}~~ \pi_i^{\text{mV}}$, and consider  three different data 
types for the response variable. The mean squared errors (MSE) and calculation time of the regression parameter 
estimators for three sampling 
probabilities are presented for comparison. 

\begin{itemize}
  \item Case 1 (continuous data): we consider a linear regression model for the longitudinal data:
$$
Y_{i}=X_{i}\beta+\epsilon_{i},~~ i=1,\cdots, m,
$$
 where the covariate $X_i$ is generated from a multivariate normal distribution  $ \mathcal{N}(0,\Sigma)$, in which the $(k,j)$ element of $\Sigma$ is $\Sigma_{kj}=0.5^{|k-j|}$.
 The regression parameter vector $\beta^{\rm T}=(0.5,0.5,1,1,2,2,10,10)$.
Two different distributions are considered for the random error vector  $\epsilon_i $: a multivariate normal distribution $\mathcal{N}(0, R_T(\alpha))$ and a multivariate Student's t-distribution with three degrees of freedom $T_3(0,R_T(\alpha))$. We consider three different correlation matrices  for $R_T(\alpha)$: the independent  matrix $I_n$
and AR(1) correlation matrix with $\alpha=0.5$.

\end{itemize}

\begin{itemize}
  \item Case 2 (count data): Correlated  count  data are generated using a log-linear regression:
  $$
  \log \mu_i =X_i\beta,
  $$
where $\beta^{\rm T}=(-0.5,-0.5,-1,1,0.5,0.5)$. The covariate $X_i$ is generated form the same distribution as  that in Case 1.

 \item Case 3 (binary data): we consider a logistic regression for the correlated binary data:
 $$
 \text{logit}  \mu_i= X_i\beta,
 $$
 where the covariate $X_i$ is generated form the same distribution as  that in Case 1, and $\beta^{\rm T}=(-0.5,-0.5,-1,1,0.5,0.5)$.
\end{itemize}

\begin{figure}[H]
  \centering  
   \subfigure[Independent]{
  \includegraphics[width=0.45\textwidth]{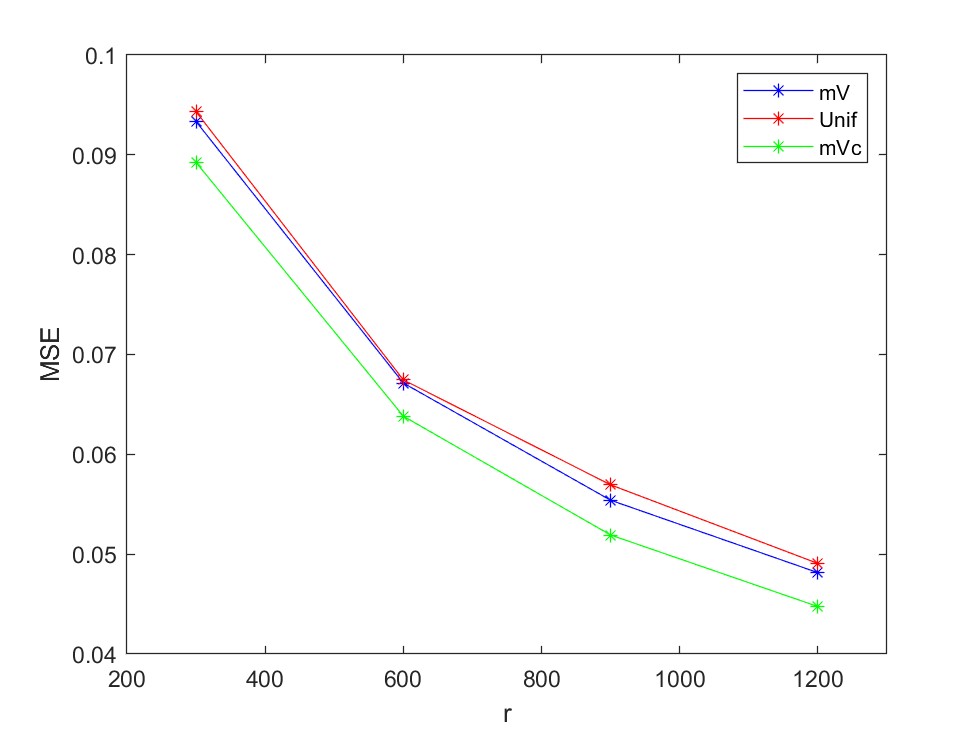}}
  \subfigure[Exchangeable]{
  \includegraphics[width=0.45\textwidth]{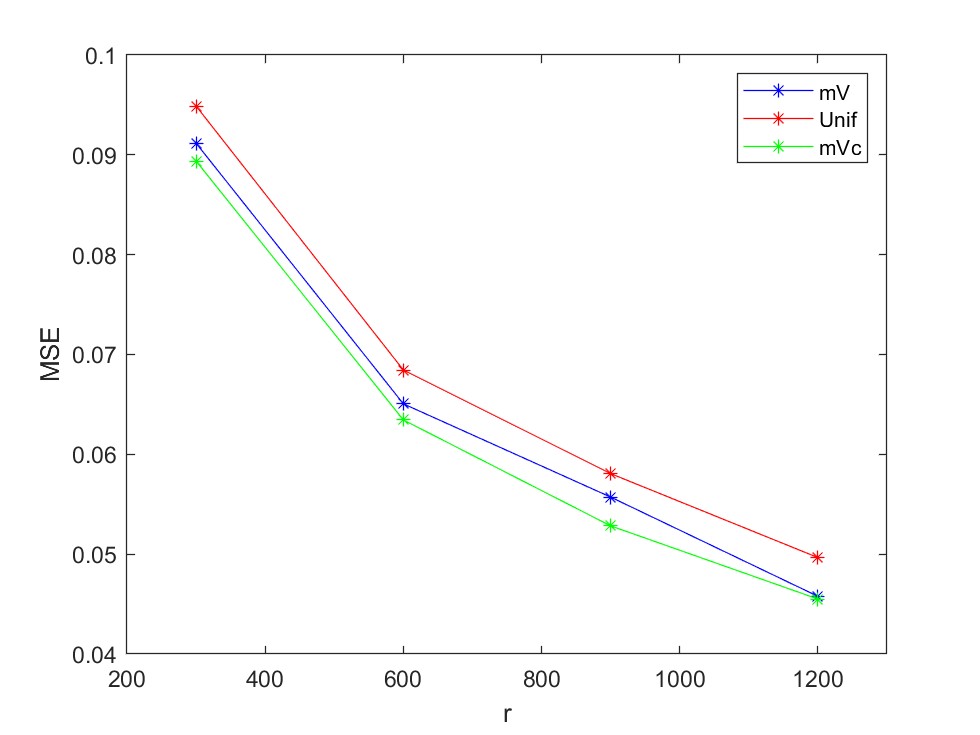}}
  \subfigure[AR(1)]{
  \includegraphics[width=0.45\textwidth]{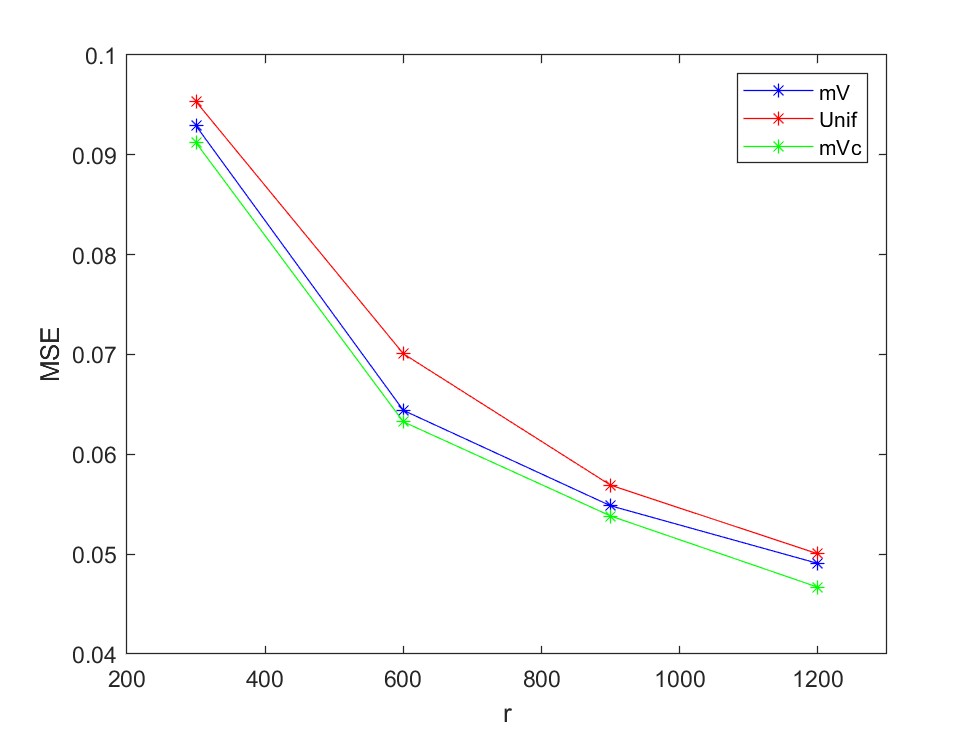}}
  \subfigure[Unstructured]{
  \includegraphics[width=0.45\textwidth]{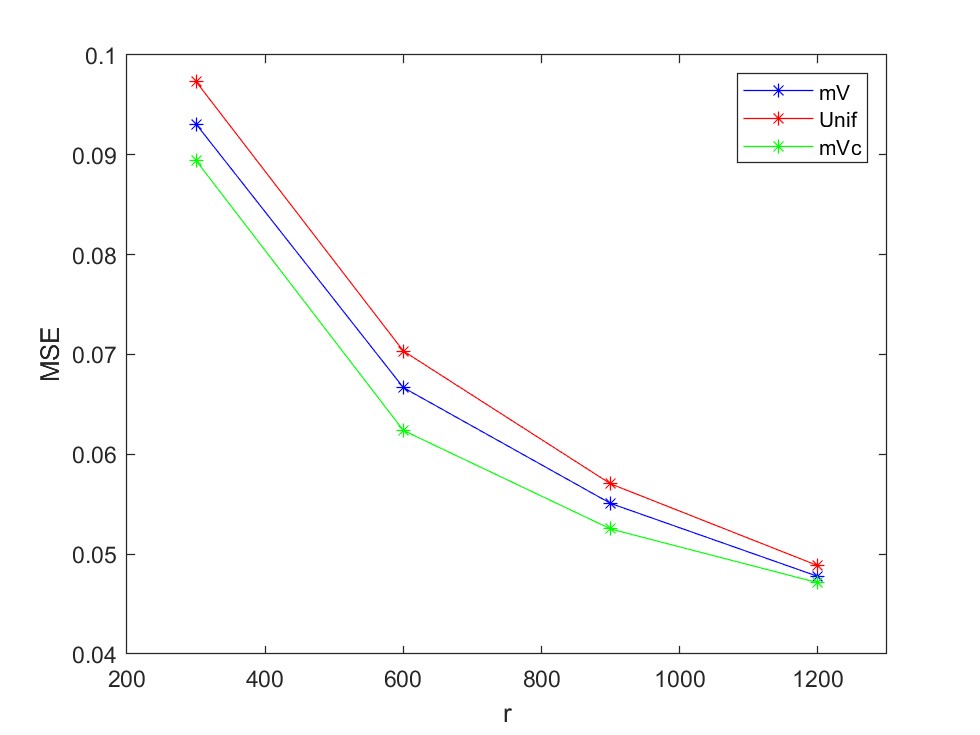}}
  \caption{The MSEs of Case 1 for the multivariate normal distribution with an independent  matrix.}
  \label{Fig.main-1}
\end{figure}

\begin{figure}[H]
  \centering  
  \subfigure[Independent]{
  \includegraphics[width=0.45\textwidth]{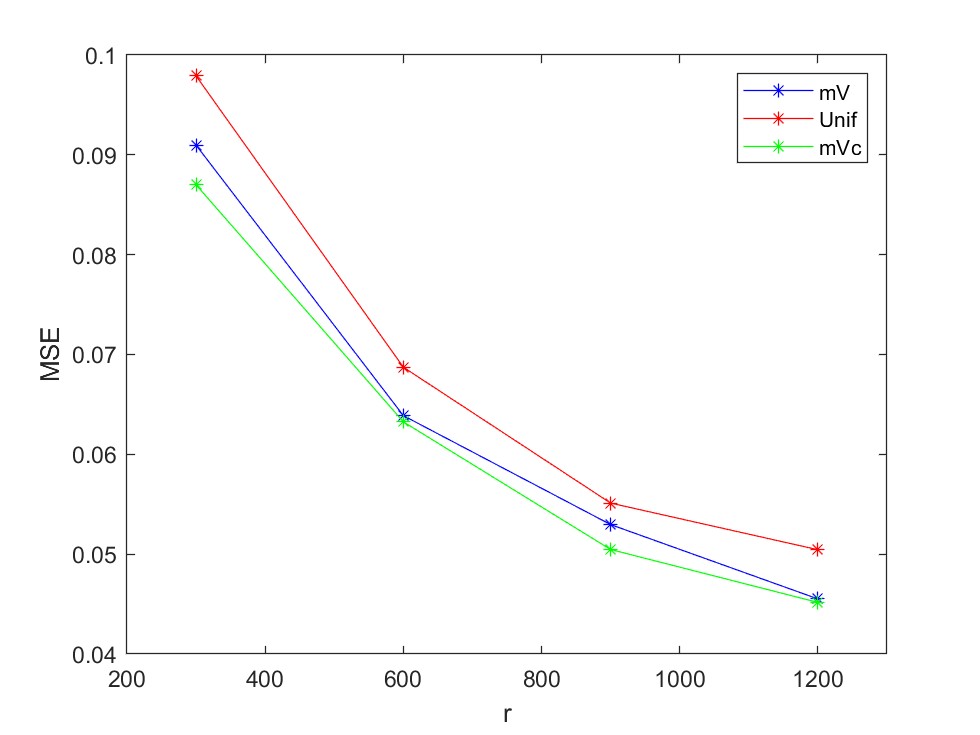}}
  \subfigure[Exchangeable]{
  \includegraphics[width=0.45\textwidth]{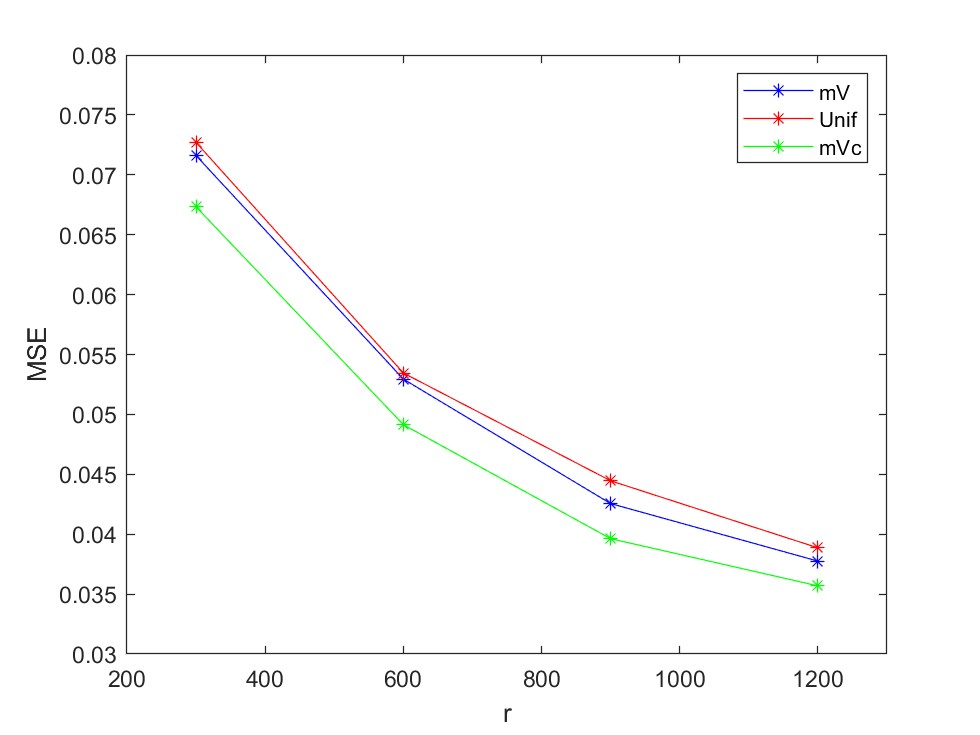}}
  \centering  
  \subfigure[AR(1)]{
  \includegraphics[width=0.45\textwidth]{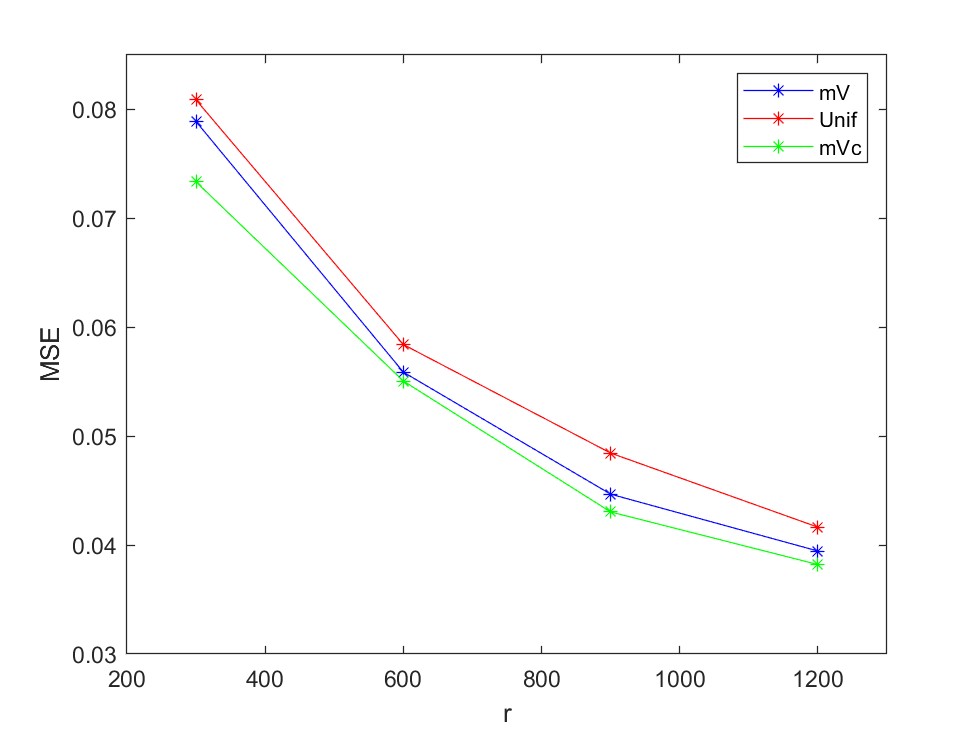}}
   \subfigure[Unstructured]{
  \includegraphics[width=0.45\textwidth]{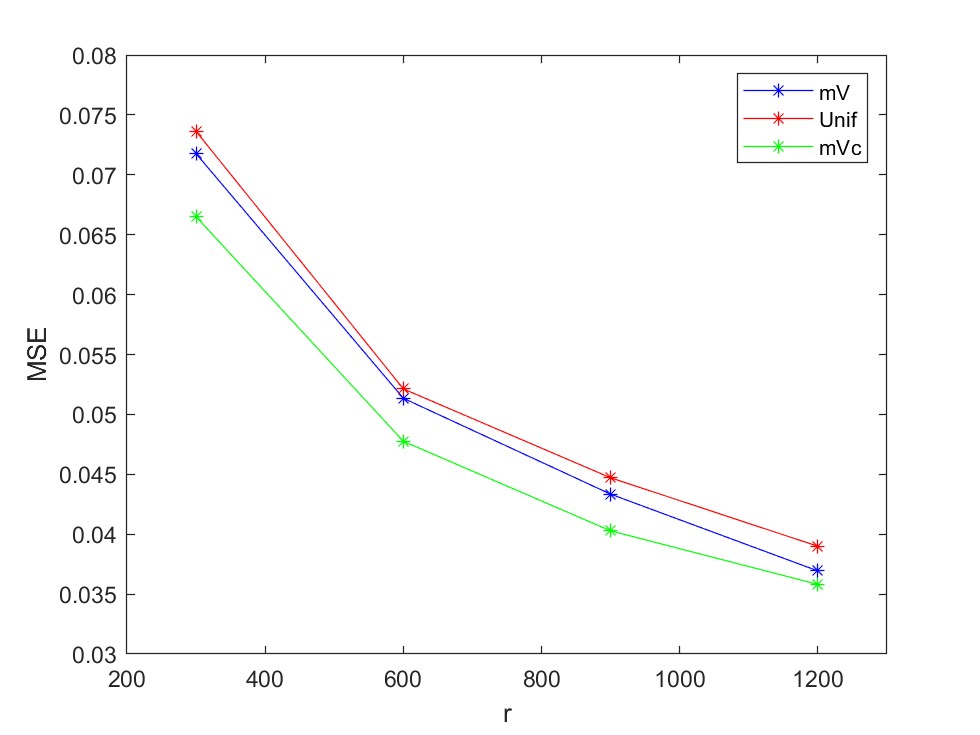}}
  \caption{ The MSEs of Case 1 for the multivariate normal distribution with an AR(1) correlation matrix.}
  \label{Fig.main-2}
\end{figure}

\begin{figure}[H]
  \centering  
  \subfigure[Independent]{

  \includegraphics[width=0.45\textwidth]{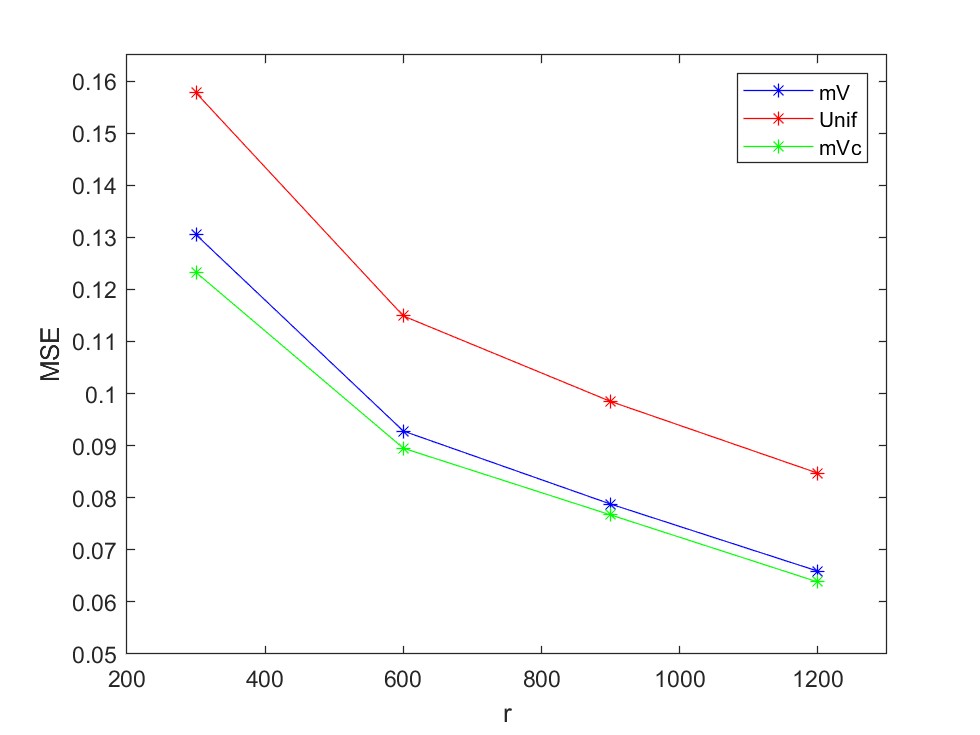}}
  \subfigure[Exchangeable]{
  \includegraphics[width=0.45\textwidth]{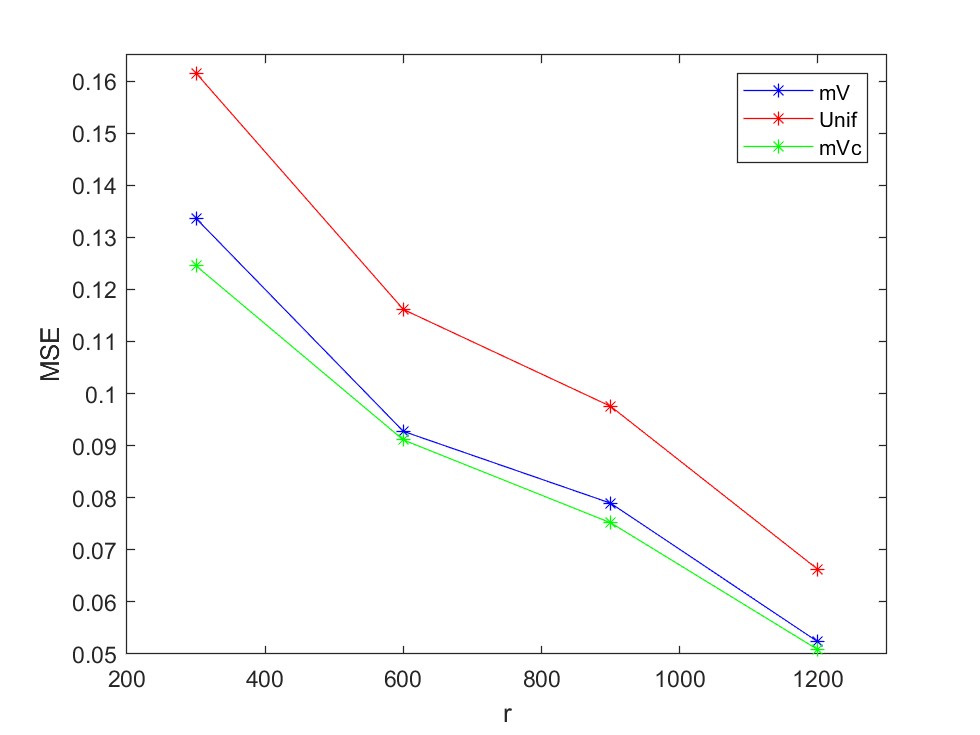}}
  \subfigure[AR(1)]{
  \includegraphics[width=0.45\textwidth]{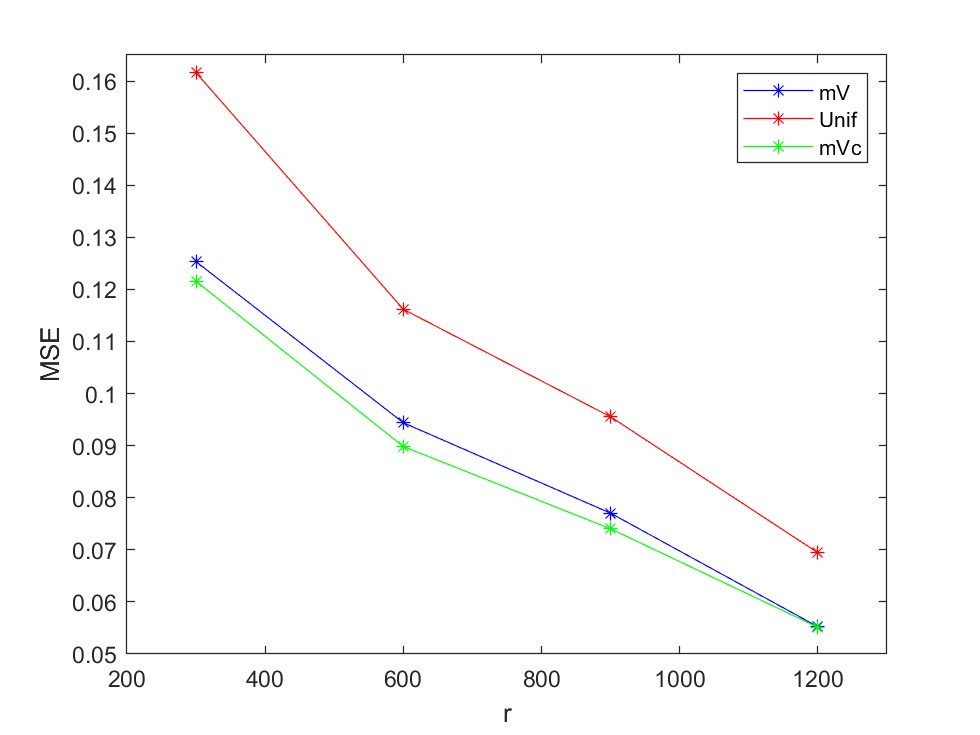}}
  \centering  
  \subfigure[Unstructured]{
  \includegraphics[width=0.45\textwidth]{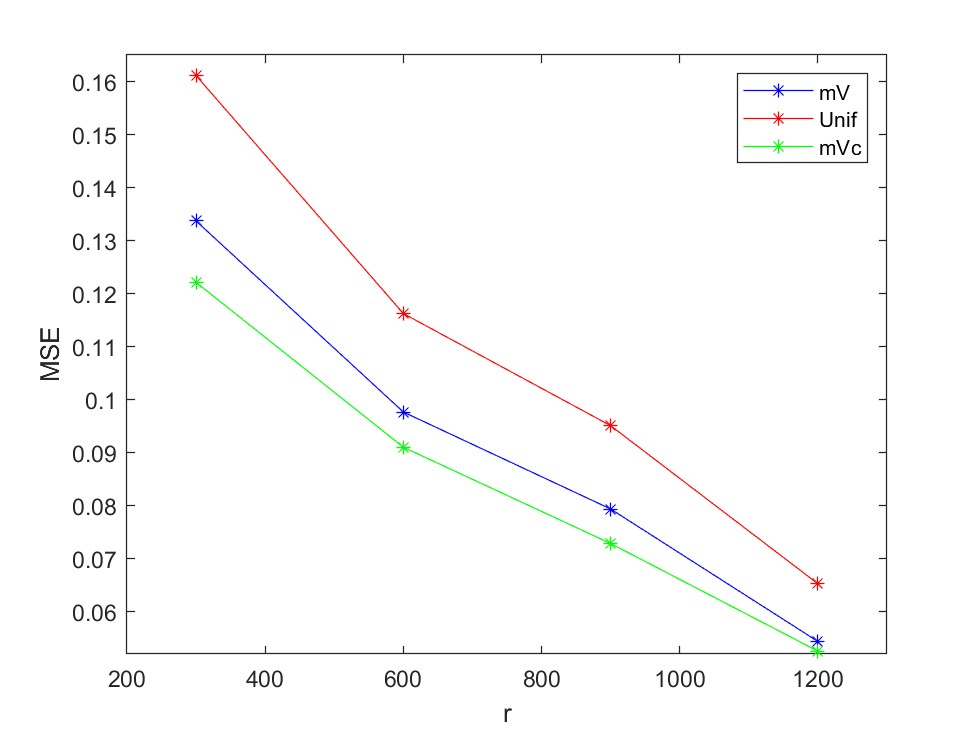}}
  \caption{The MSEs of Case 1 for the multivariate t distribution with an independent  matrix.}
  \label{Fig.main-3}
\end{figure}

\begin{figure}[H]
  \centering  
  \subfigure[Independent]{
  \includegraphics[width=0.45\textwidth]{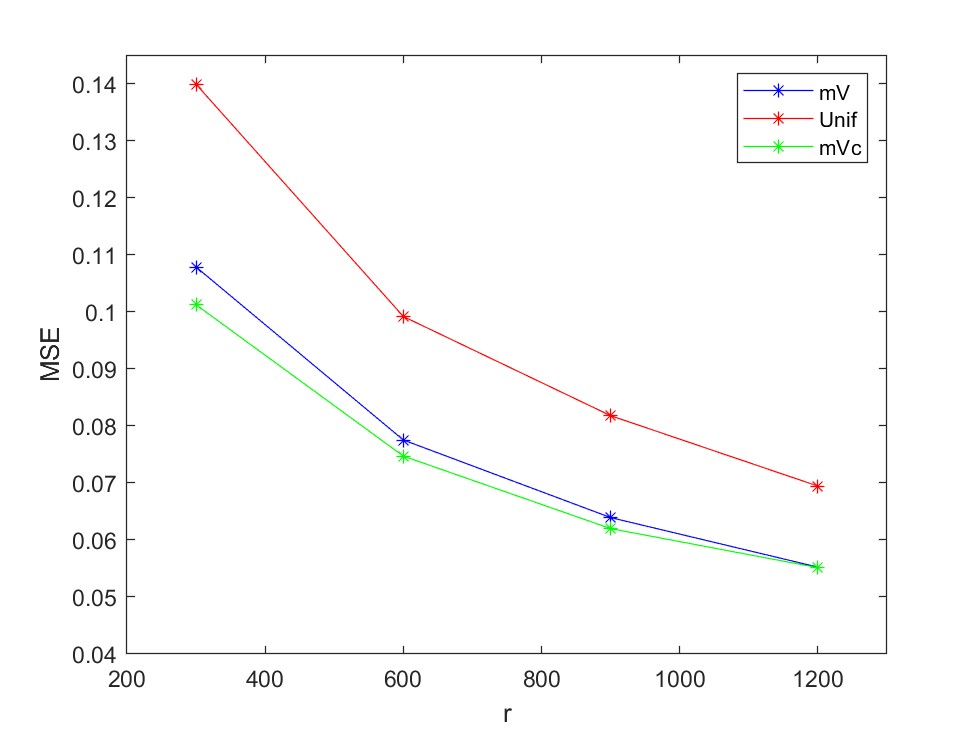}}
  \subfigure[Exchangeable]{
  \includegraphics[width=0.45\textwidth]{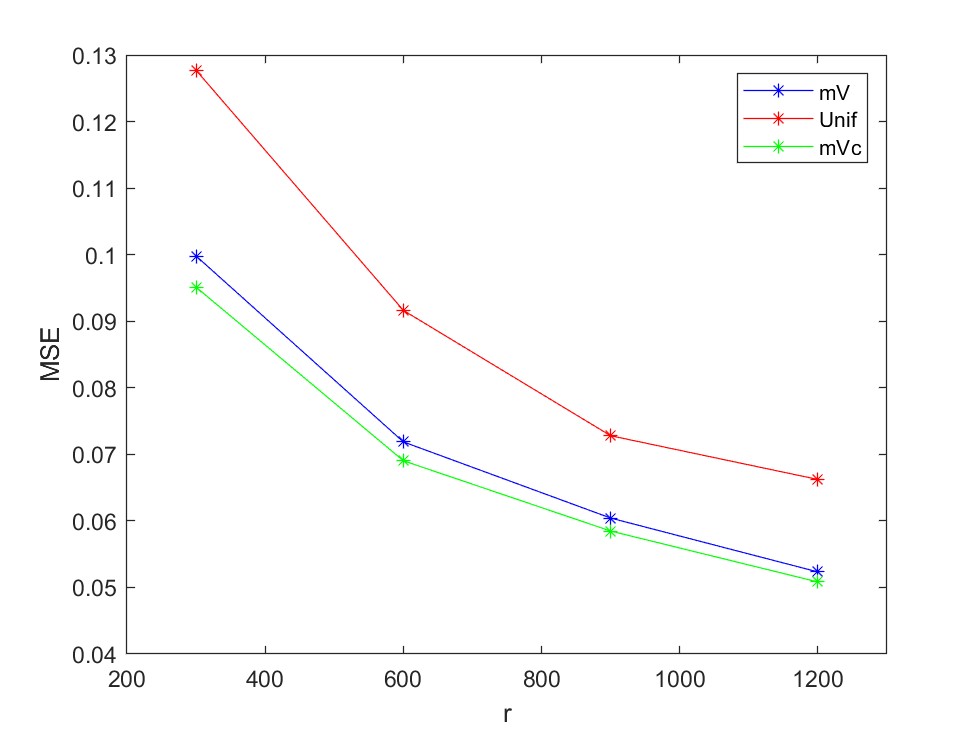}}
  \centering  
  \subfigure[AR(1)]{
  \includegraphics[width=0.45\textwidth]{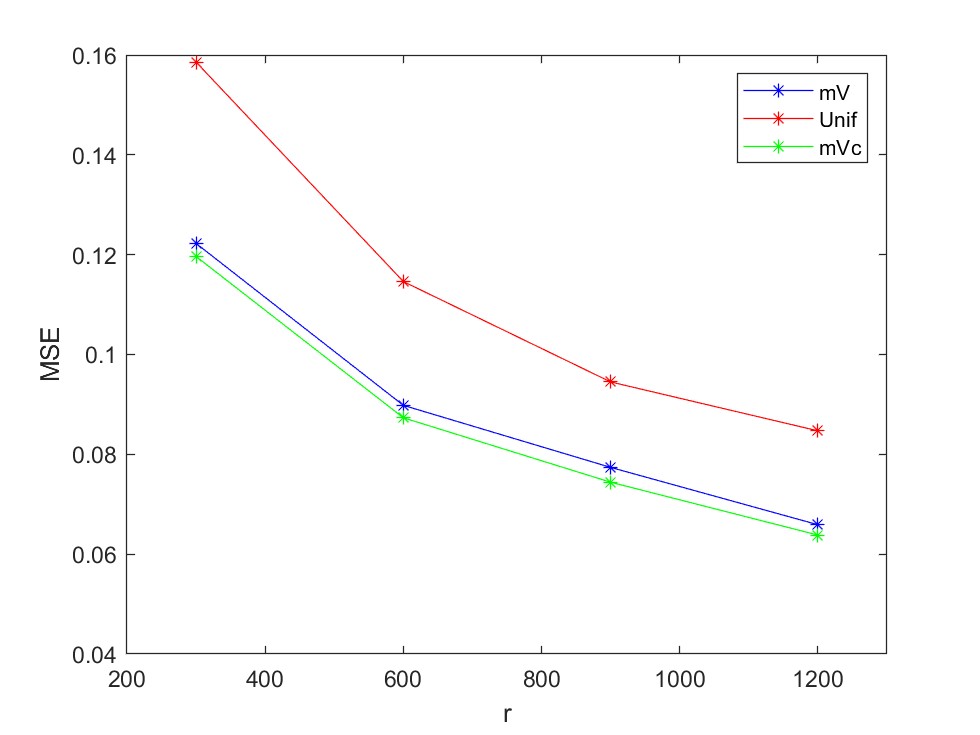}}
  \subfigure[Unstructured]{
  \includegraphics[width=0.45\textwidth]{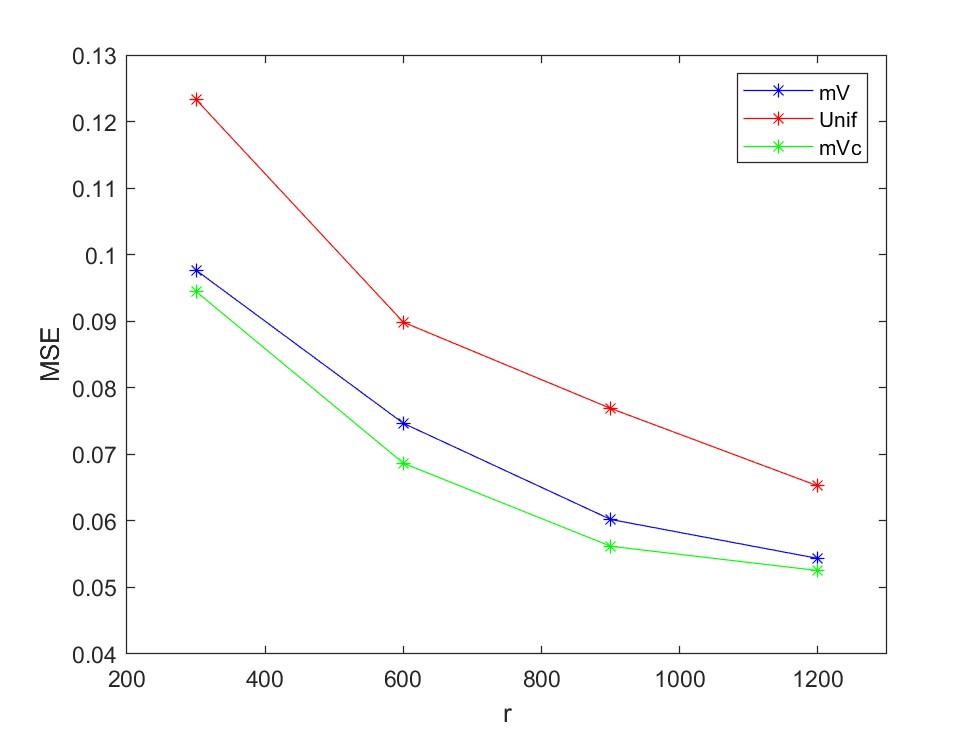}}
  \caption{The MSEs of Case 1 for the multivariate t distribution with an AR(1) correlation matrix.}
  \label{Fig.main-4}
\end{figure}

\begin{figure}[H]
  \centering  
  \subfigure[Independent]{
  \includegraphics[width=0.45\textwidth]{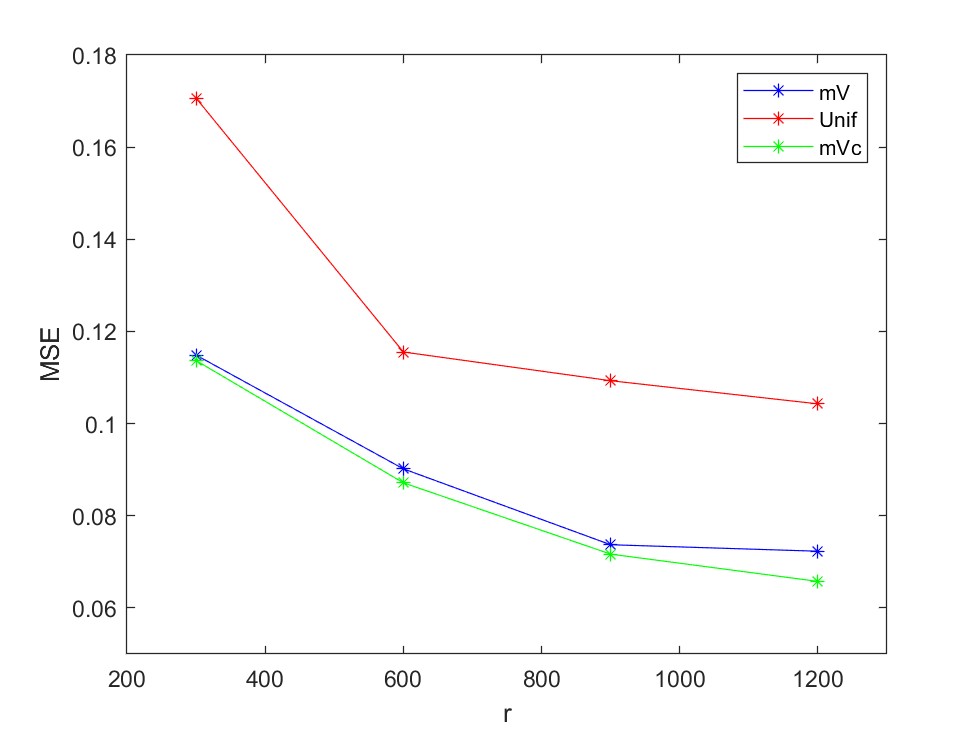}}
  \subfigure[Exchangeable]{
  \includegraphics[width=0.45\textwidth]{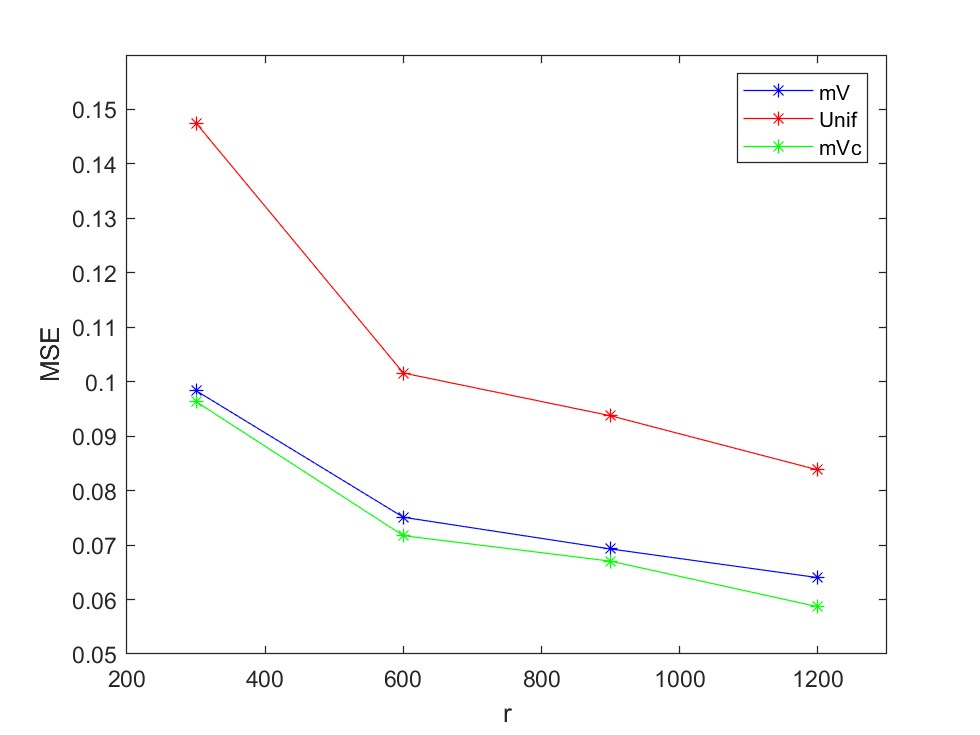}}
  \centering  
  \subfigure[AR(1)]{
  \includegraphics[width=0.45\textwidth]{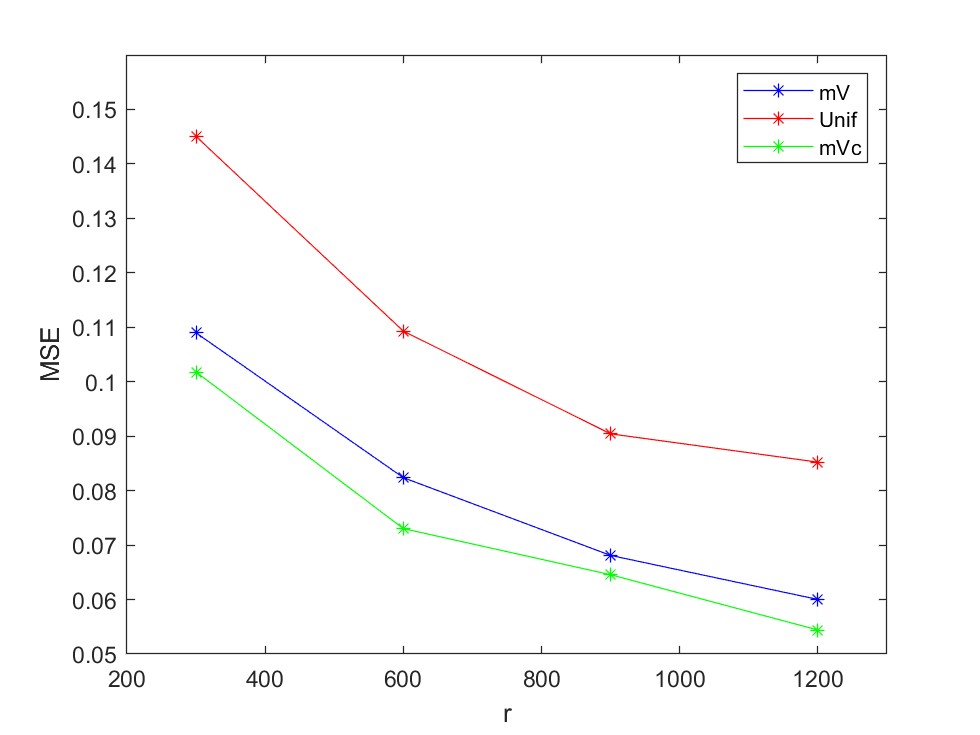}}
  \subfigure[Unstructured]{
  \includegraphics[width=0.45\textwidth]{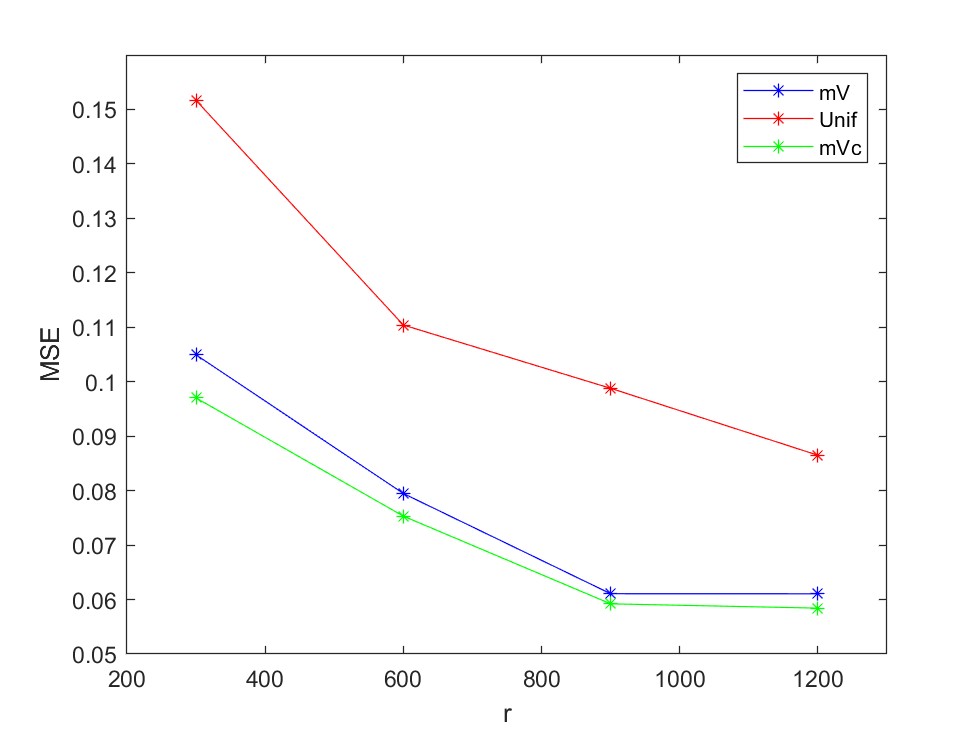}}
  \caption{The MSEs of Case 2 with an independent  matrix.}
  \label{Fig.main-5}
\end{figure}

\begin{figure}[H]
  \centering  
  \subfigure[Independent]{
  \includegraphics[width=0.45\textwidth]{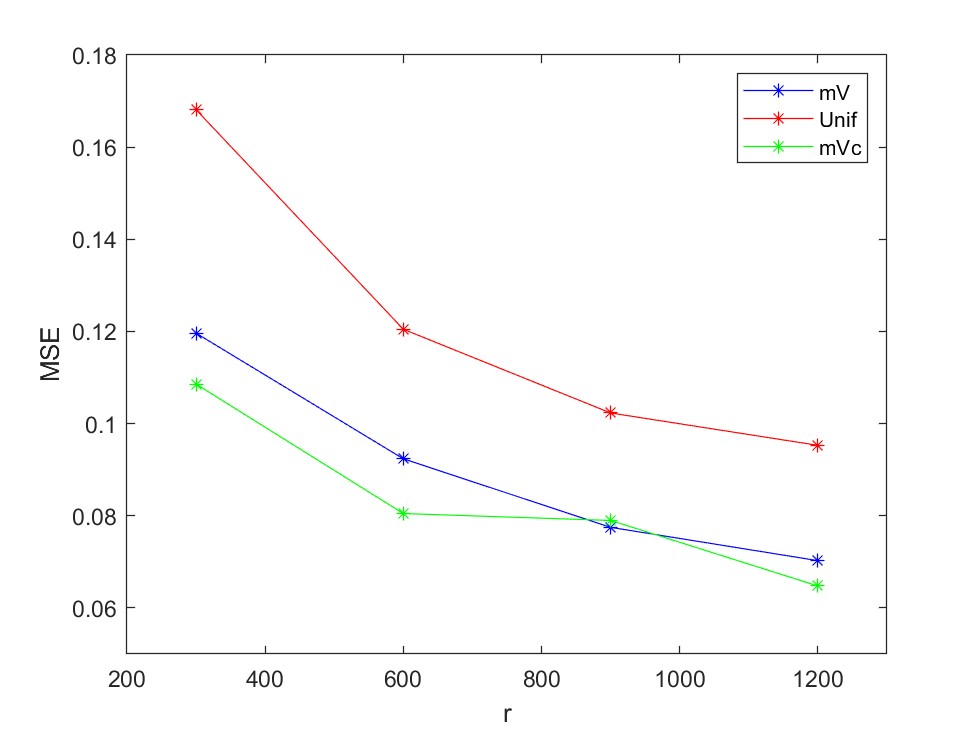}}
  \subfigure[Exchangeable]{
  \includegraphics[width=0.45\textwidth]{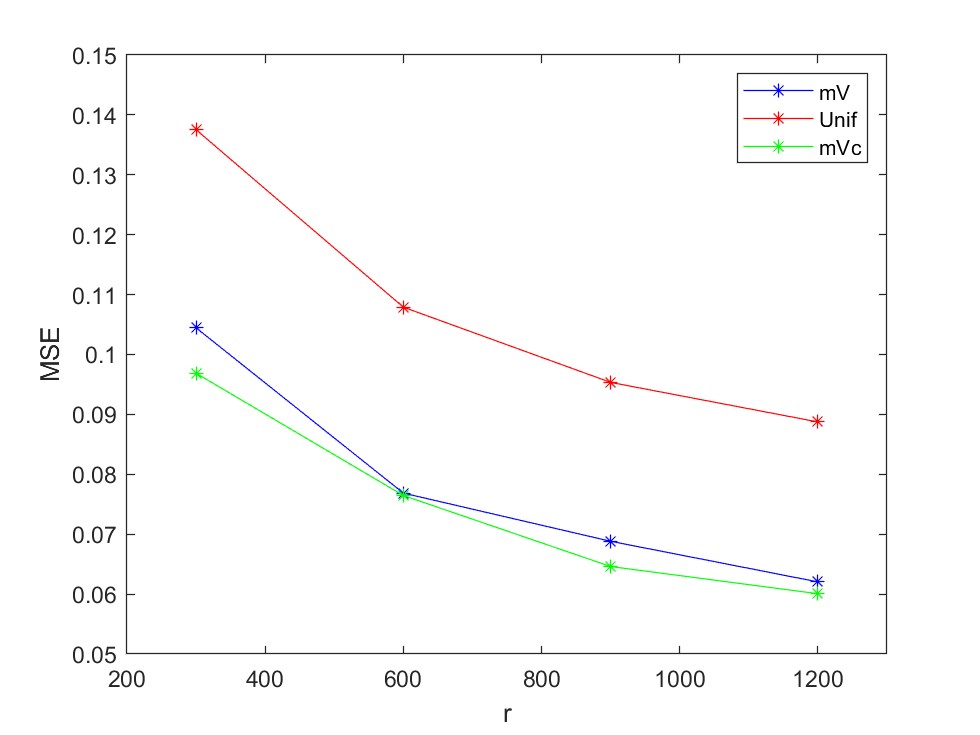}}
  \centering  
  \subfigure[AR(1)]{
  \includegraphics[width=0.45\textwidth]{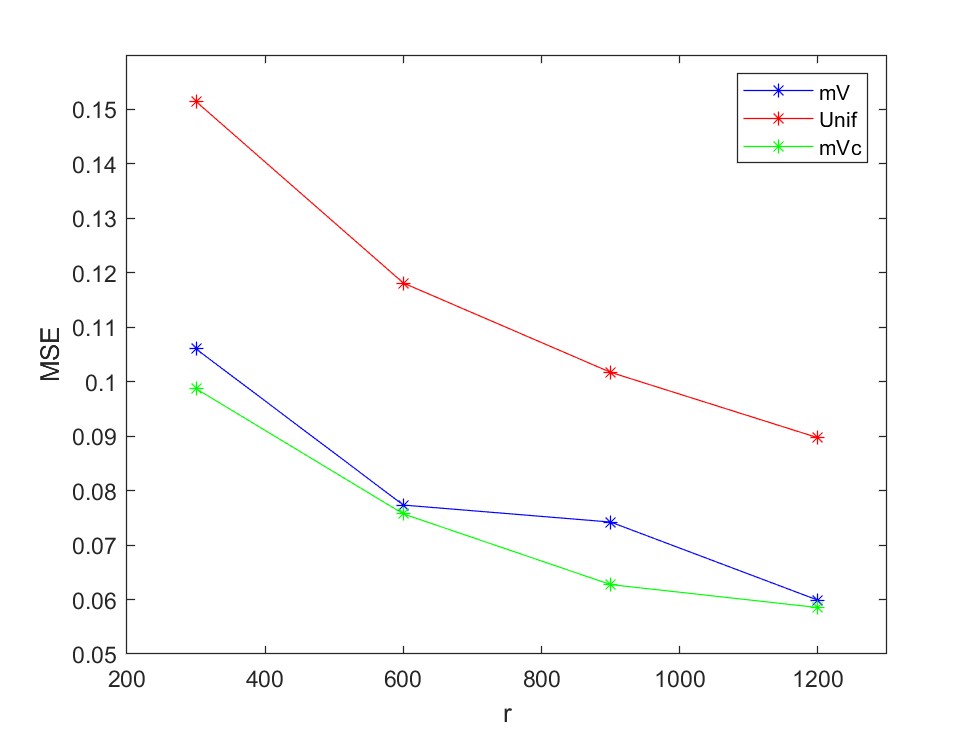}}
  \subfigure[Unstructured]{
  \includegraphics[width=0.45\textwidth]{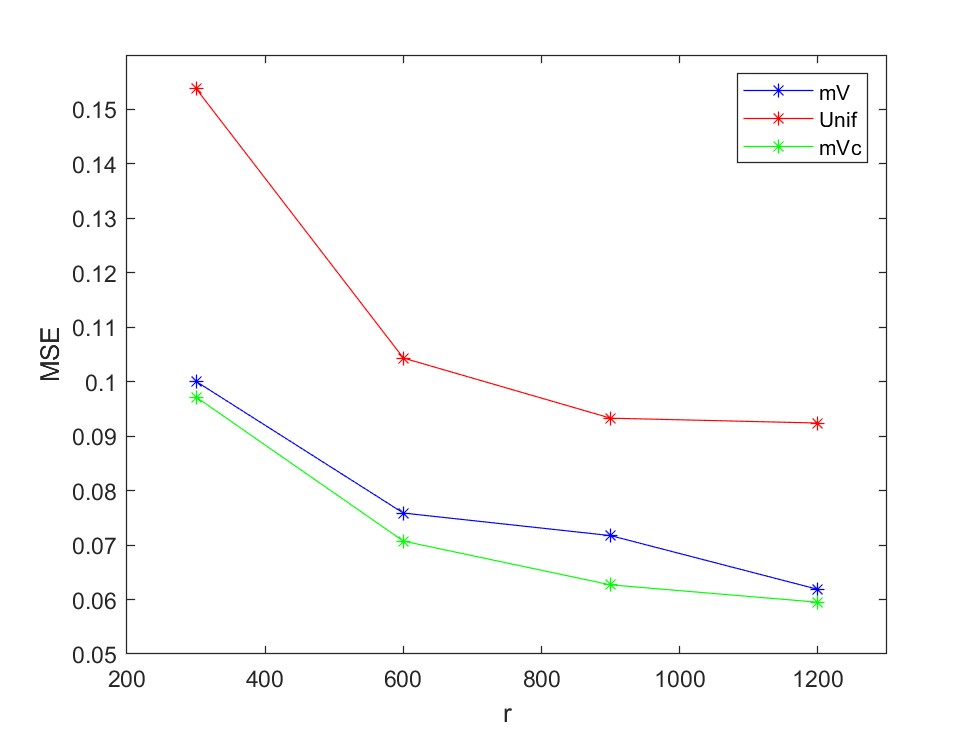}}
  \caption{The MSEs of Case 2 with an AR(1) correlation matrix.}
  \label{Fig.main-6}
\end{figure}

\begin{figure}[H]
  \centering  
  \subfigure[Independent]{
  \includegraphics[width=0.45\textwidth]{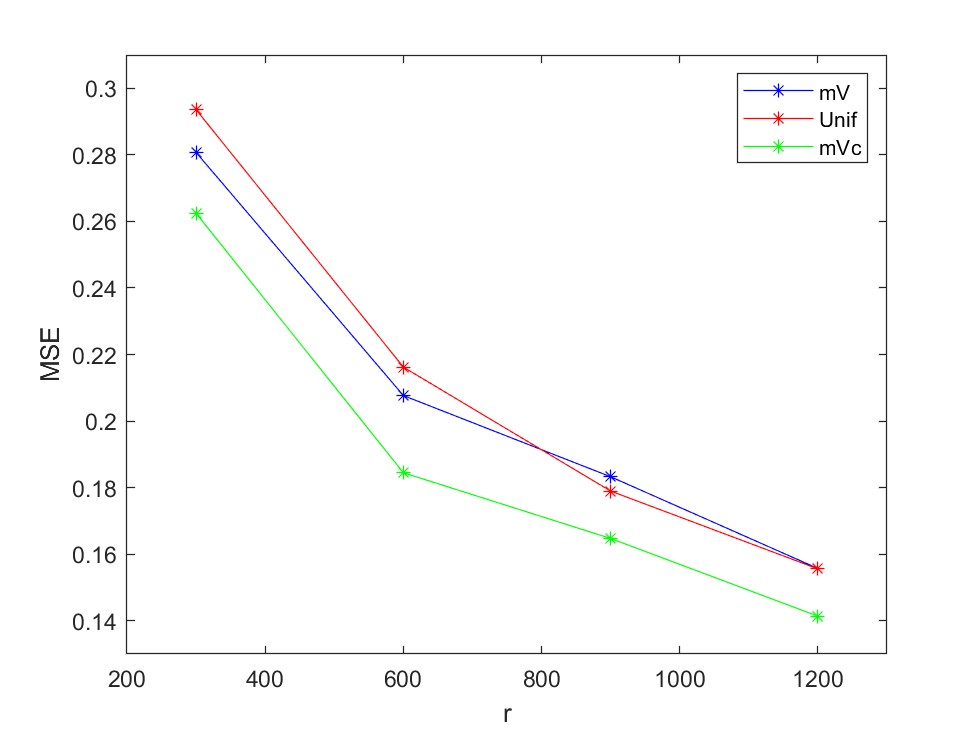}}
  \subfigure[Exchangeable]{
  \includegraphics[width=0.45\textwidth]{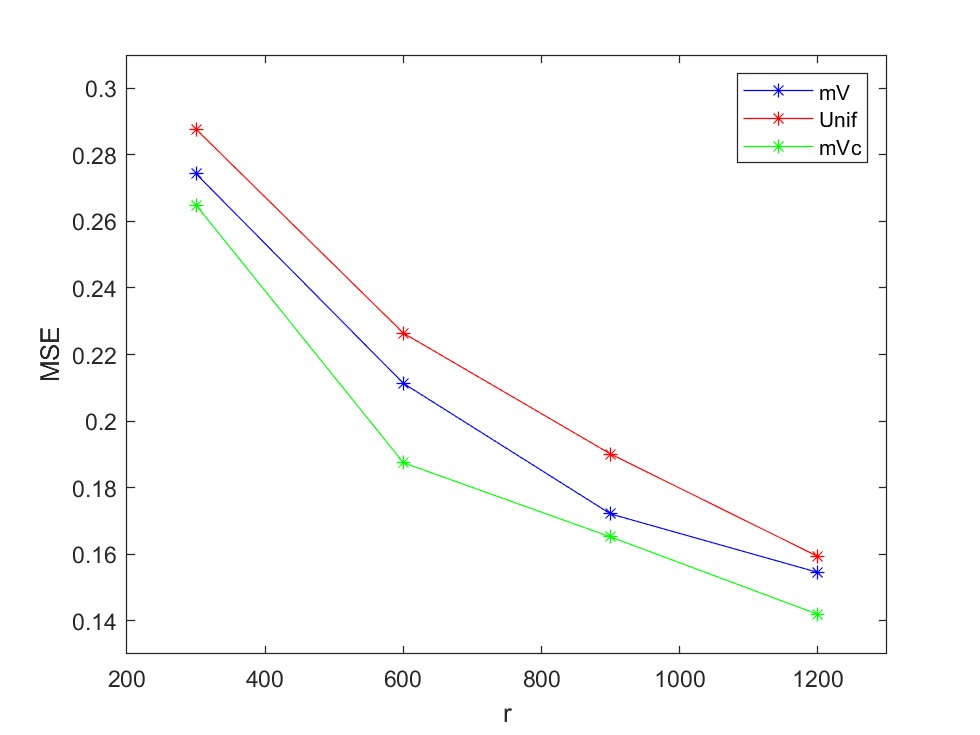}}
  \centering  
  \subfigure[AR(1)]{
  \includegraphics[width=0.45\textwidth]{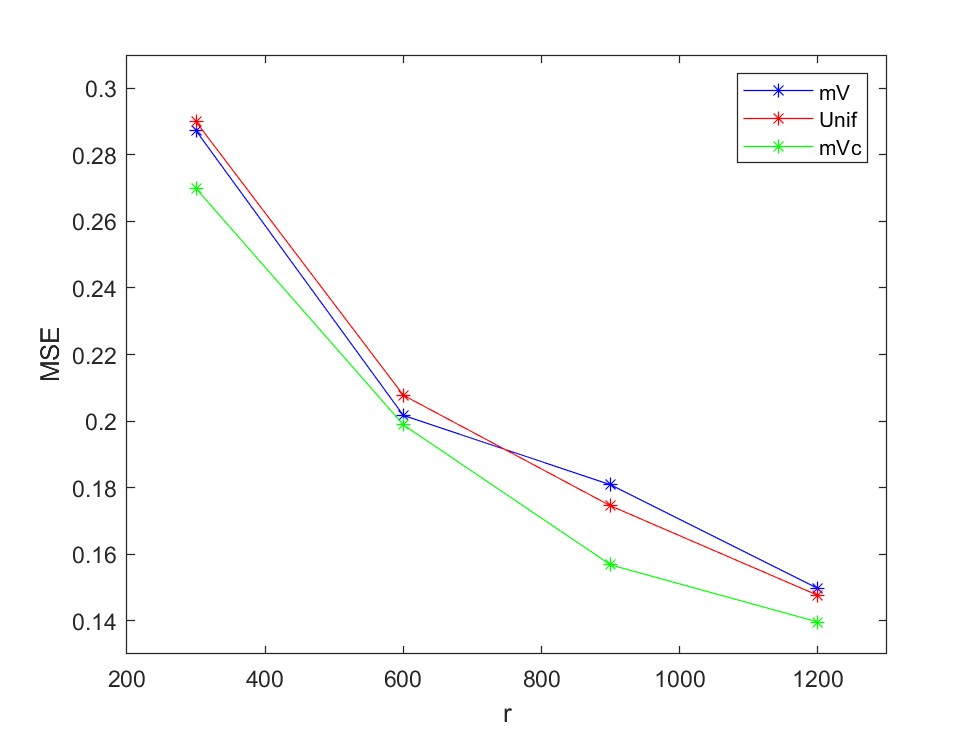}}
  \subfigure[Unstructured]{
  \includegraphics[width=0.45\textwidth]{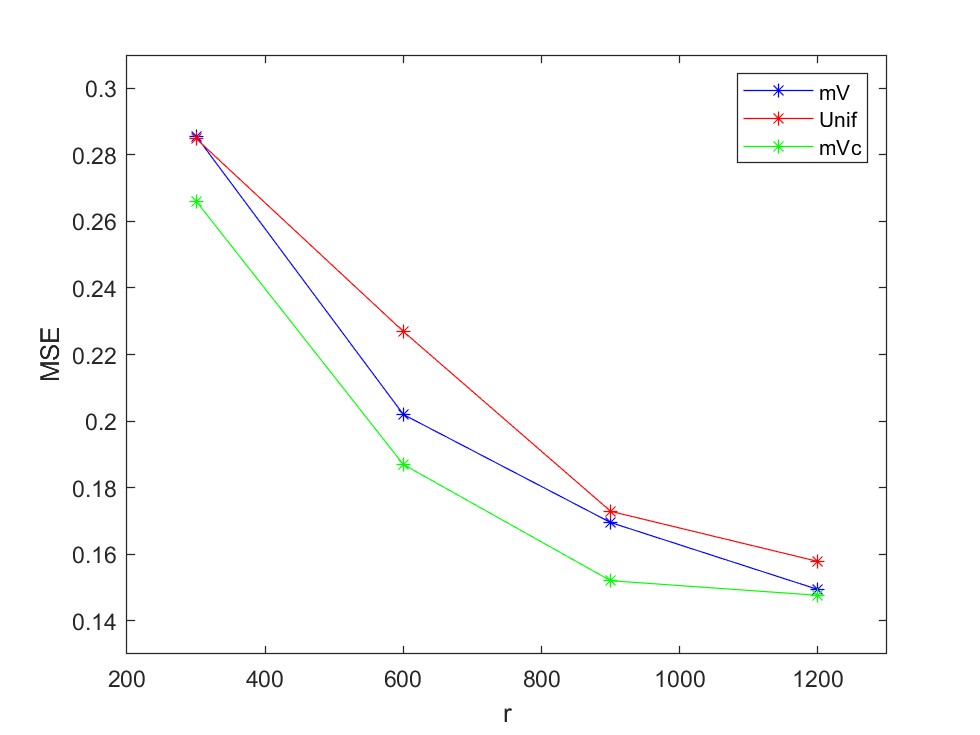}}
  \caption{The MSEs of Case 3 with an independent  matrix.}
  \label{Fig.main-7}
\end{figure}

\begin{figure}[H]
  \centering  
  \subfigure[Independent]{
  \includegraphics[width=0.45\textwidth]{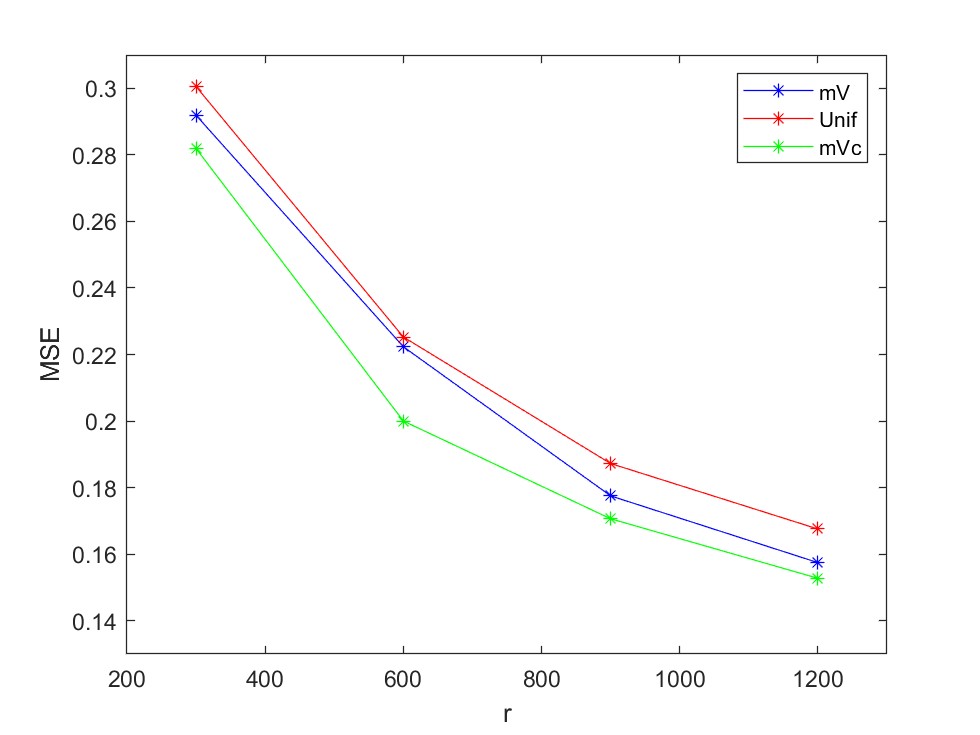}}
  \subfigure[Exchangeable]{
  \includegraphics[width=0.45\textwidth]{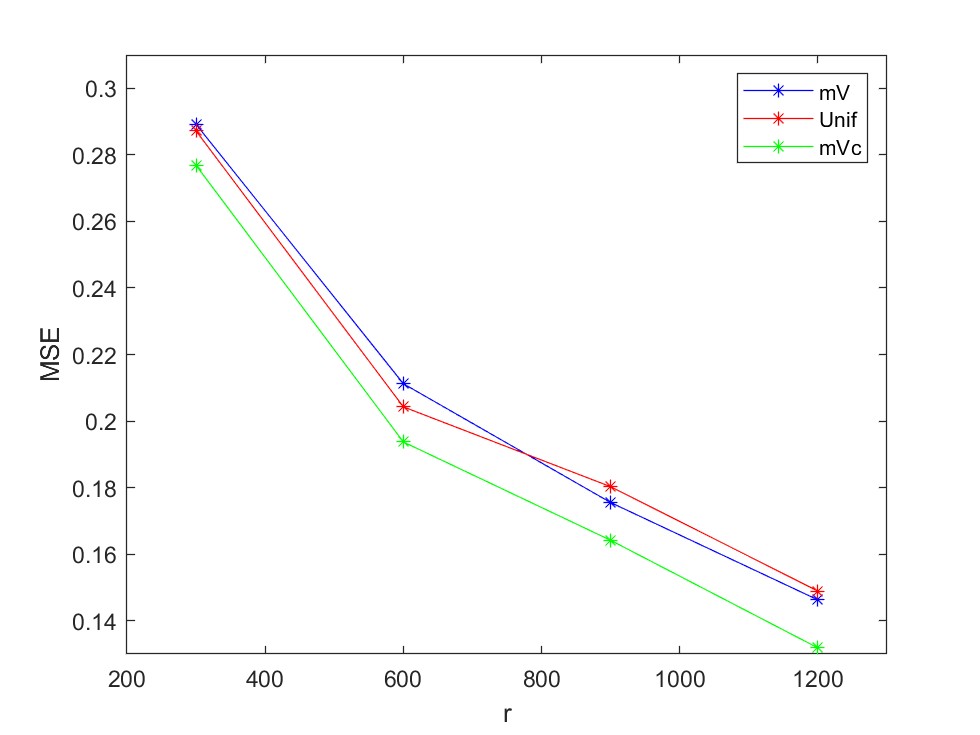}}
  \centering  
  \subfigure[AR(1)]{
  \includegraphics[width=0.45\textwidth]{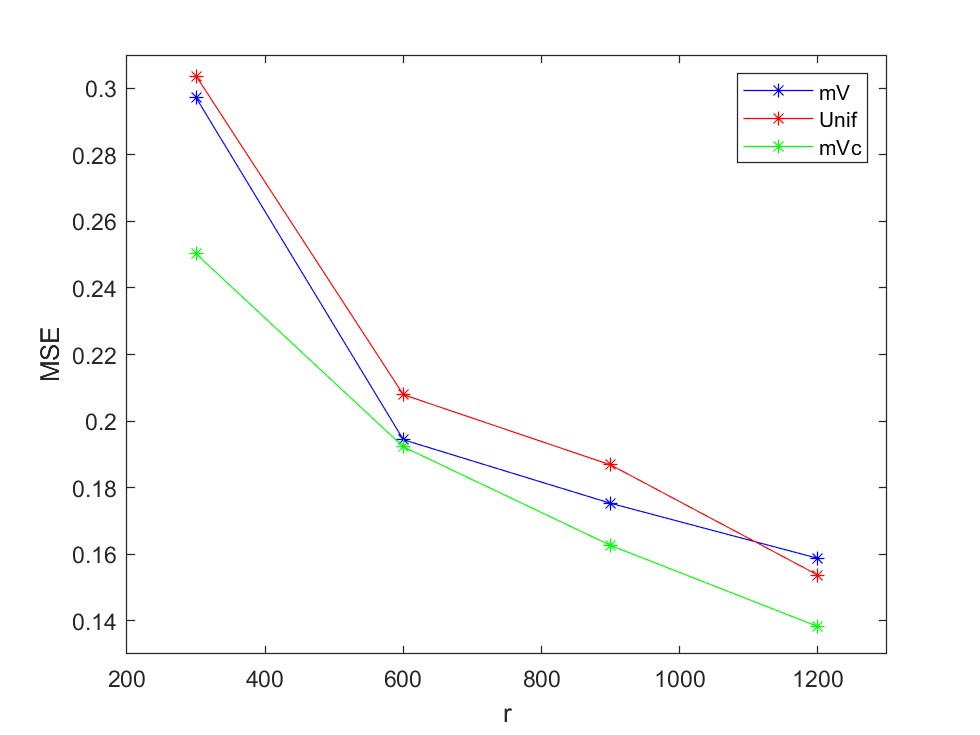}}
  \subfigure[Unstructured]{
  \includegraphics[width=0.45\textwidth]{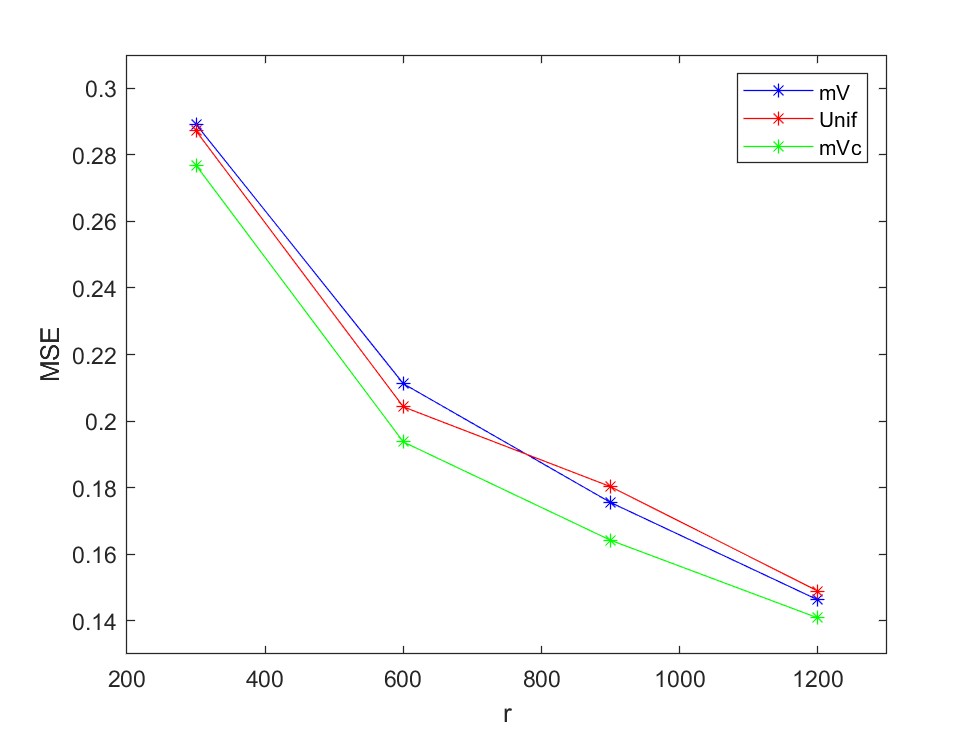}}
  \caption{The MSEs of Case 3 with an AR(1) correlation matrix.}
  \label{Fig.main-8}
\end{figure}

In Case 2 and Case 3, the setup for the correlation matrix of $Y_i$ is the same as that in Case 1.
 For each case, the observation times for each  subject is set as $n=5$. The  sample size $m$ is $10000$.
The size of subsample $r$ is $300$, $600$, $900$, $1200$, and $r_0=200$. The working correlation matrix structure
is considered as identity, exchangeable, AR(1) and unstructured.
In each simulation study, $200$ realizations are carried out. 
The MSEs of Cases 1-3
with different working correlation matrices are presented in Figures \ref{Fig.main-1}--\ref{Fig.main-8}. 
The average calculating time for three cases within an AR(1) correlation matrix are presented in Tables \ref{tab:case1}-- \ref{tab:case3}. 

According to Figures \ref{Fig.main-1}--\ref{Fig.main-4}, for the continuous data, the MSEs decrease as the number of subsampling samples increases. 
Both mVc and mV are significantly better than
the uniform sampling probability for different error distributions and different working correlation matrices. 
The mVc  always performs better than the other two sampling probabilities.
The difference between the MSEs of mV and the uniform sampling probability is sometimes not significant. The proposed methods still perform well even when the working correlation structure is misspecified.
 For the count data, the results are presented in Figures  \ref{Fig.main-5}--\ref{Fig.main-6}.
 It can be seen that the results have the same pattern as those for the continuous data. The MSEs of mV and mVc are smaller than those of the uniform sampling probability. The  mV probability  is sometimes comparable with the mVc probability.
For the binary data (Figures  \ref{Fig.main-7}--\ref{Fig.main-8}),
the results have the similar pattern as those for the continuous and count data in most cases. In addition,  the uniform sampling method is sometimes comparable with the mV method.  The  mVc is still the  best in this three sampling probabilities.
 

 From Tables \ref{tab:case1}--\ref{tab:case3}, we can see that the calculate time increases as the sampling size increases.
 The calculate time with the sampling subsets is shorter than that using the full data especially for the logistic regression. For the linear regression with the longitudinal data, the calculate time of mVc is comparable with that of mV. The computation speed based on the uniform probabilities is the fastest. For the log-linear and logistic regression models, the calculation time based on the uniform and mV probabilities is shorter than 
 that based on the mVc probabilities in most cases. 


  \begin{table}[H]
    \caption{The average calculating time of Case 1 for the multivariate normal distribution with an AR(1) correlation matrix.}\label{tab:case1}
    \centering
    \begin{tabular}{@{}llrrrr@{}}
    \toprule
    \multicolumn{2}{c}{\multirow{2}{*}{}} & \multicolumn{4}{c}{Working correlation matrix}                                                                                                   \\
    \multicolumn{2}{c}{}                  & \multicolumn{1}{c}{Independent} & \multicolumn{1}{c}{AR(1)} & \multicolumn{1}{c}{Exchangeable} & \multicolumn{1}{c}{Unstructured} \\ \midrule
    \multirow{3}{*}{$r=300$}  & Uniform   & 0.058                           & 0.061                     & 0.056                           & 0.071                           \\
                              & mV        & 0.077                           & 0.082                     & 0.074                           & 0.071                           \\
                              & mVc       & 0.078                           & 0.089                     & 0.076                           & 0.075                           \\
      \ \ \\

\multirow{3}{*}{$r=600$}  & Uniform   & 0.085                           & 0.081                  & 0.086                           & 0.097                          \\
 & mV        & 0.086                           & 0.095                     & 0.094                           & 0.101        \\
                              & mVc       & 0.094                           & 0.094                     & 0.096                           & 0.099                           \\
                                 \ \  \\
    \multirow{3}{*}{$r=900$}  & Uniform   & 0.101                           & 0.104                     & 0.105                           & 0.114                           \\
                                 & mV        & 0.107                           & 0.110                     & 0.108                           & 0.116                           \\
                                 & mVc       & 0.108                           & 0.114                     & 0.108                           & 0.120                           \\
                                     \ \ \\

    \multirow{3}{*}{$r=1200$} & Uniform   & 0.120                           & 0.124                     & 0.121                           & 0.133                           \\
                              & mV        & 0.122                           & 0.120                     & 0.128                           & 0.139                           \\
                              & mVc       & 0.131                           & 0.127                     & 0.130                           & 0.139                           \\
                              &Full data  & 0.919                          & 0.912                    & 0.924                          &  0.969                         \\ \bottomrule
    \end{tabular}
    \end{table}

    \begin{table}[H]
        \caption{The average calculating time of Case 2 with an AR(1) correlation matrix.}\label{tab:case2}
        \centering
        \begin{tabular}{@{}llrrrr@{}}
        \toprule
        \multicolumn{2}{c}{\multirow{2}{*}{}} & \multicolumn{4}{c}{Working correlation matrix}                                                                                                   \\
    \multicolumn{2}{c}{}                  & \multicolumn{1}{c}{Independent} & \multicolumn{1}{c}{AR(1)} & \multicolumn{1}{c}{Exchangeable} & \multicolumn{1}{c}{Unstructured} \\ \midrule
        \multirow{3}{*}{$r=300$}  & Uniform   & 0.626                           & 0.574                     & 0.661                           & 0.695                           \\
                                  & mV        & 0.696                           & 0.670                     & 0.699                           & 0.721                           \\
                                  & mVc       & 0.698                           & 0.747                     & 0.746                           & 0.769                           \\
          \ \ \\
    
\multirow{3}{*}{$r=600$}  & Uniform   & 0.888                           & 0.873                     & 0.862                           & 0.913                           \\
                                  & mV        & 0.904                           & 0.949                     & 0.923                           & 0.941                           \\
                                  & mVc       & 0.971                           & 0.955                     & 0.971                           & 0.980                           \\
                                     \ \  \\
        \multirow{3}{*}{$r=900$}  & Uniform   & 1.042                           & 1.014                     & 1.011                           & 1.089                           \\
                                     & mV        & 1.149                           & 1.097                     & 1.097                           & 1.161                           \\
                                     & mVc       & 1.168                           & 1.143                     & 1.112                           & 1.201                           \\
                                         \ \ \\

        \multirow{3}{*}{$r=1200$} & Uniform   &1.500	&1.495	&1.464	&1.481        \\
                                  & mV        & 1.538	&1.535	&1.524	&1.503                                  \\
                                  & mVc       & 1.677	&1.687	&1.673	&1.707                                  \\
                                  &Full data  & 57.411           & 59.478                    & 58.015                          & 60.137                          \\ \bottomrule
        \end{tabular}
        \end{table}
        
        \begin{table}[H]
          \caption{The average calculating time of Case 3 with an AR(1) correlation matrix.}\label{tab:case3}
          \centering
          \begin{tabular}{@{}llrrrr@{}}
          \toprule
          \multicolumn{2}{c}{\multirow{2}{*}{}} & \multicolumn{4}{c}{Working correlation matrix}                                                                                                   \\
          \multicolumn{2}{c}{}                  & \multicolumn{1}{c}{Independent} & \multicolumn{1}{c}{AR(1)} & \multicolumn{1}{c}{Exchangeable} & \multicolumn{1}{c}{Unstructured} \\ \midrule
          \multirow{3}{*}{$r=300$}  & Uniform   & 0.326                           & 0.311                     & 0.326                           & 0.390                           \\
          & mV        & 0.314                           & 0.314                     & 0.334                           & 0.533                           \\
          & mVc       & 0.375                           & 0.370                     & 0.367                           & 0.363                           \\
            \ \ \\
      
          \multirow{3}{*}{$r=600$}  & Uniform   & 0.488                           & 0.451                     & 0.435                           & 0.413                           \\
                                    & mV        & 0.494                           & 0.478                     & 0.406                           & 0.446                           \\
                                    & mVc       & 0.555                           & 0.555                     & 0.521                           & 0.480                           \\
                                       \ \  \\
                                       \multirow{3}{*}{$r=900$}  & Uniform   & 0.523                           & 0.548                     & 0.520                           & 0.579                           \\
                                       & mV        & 0.517                           & 0.527                     & 0.428                           & 0.566                           \\
                                       & mVc       & 0.668                           & 0.621                     & 0.645                           & 0.529                           \\
                                           \ \ \\

          \multirow{3}{*}{$r=1200$} & Uniform   & 0.684                           & 0.684                     & 0.717                           & 0.664                           \\
                                    & mV        & 0.704                           & 0.663                     & 0.686                           & 0.884                           \\
                                    & mVc       & 0.774                           & 0.758                     & 0.772                           & 0.746                           \\
                                    &Full data  & 16.542                          & 16.578                    & 18.269                          & 20.395                          \\ \bottomrule
          \end{tabular}
          \end{table}
\section{Real data analysis}
In this section, we use a depression data set to illustrate the proposed method.
The data is obtained from the CHARLS (China Health and Retirement Longitudinal Study) database (http://charls.pku.edu.cn/), which is hosted by the National Development Institute of Peking University \citep{zhao2013}.
The response variable in the depression data is depression scores, and the related factors include Age, Gender, Area, Education, Disease, Marital status, Life satisfaction, Eyesight, 
Hearing, Teeth state, Drinking, Nap time and Social activity.  Gender takes a value of 1 and
 2 for male and female, respectively; Age is represented by year; Area is represented 
 by 0 and 1 for rural and urban areas, respectively; 
 Education is represented by 1, 2 and 3 for primary school and below, junior or senior secondary school and university and 
 above, respectively; 
 Disease is represented by 0, 1 and 2 for no chronic disease, one chronic disease, two or more chronic diseases;
 marital status is represented by 1, 2 and 3 for married and living together or temporarily separated for work, separated, divorced or widowed or never married and living together;
 Life satisfaction is represented by 1 and 2 for satisfied with life and dissatisfied with life;
 Eyesight is represented by 1, 2 and 3 for good vision, fair vision and poor vision, respectively;
 Hearing is represented by 1, 2 and 3 for good hearing, fair hearing and poor hearing, respectively;
 Teeth state is represented by 1 and 2 for wearing dentures and not wearing dentures, respectively;
 Drinking is represented by 1, 2 and 3 for drinking more often than once a month, less often than once a month and not drinking alcohol respectively;
 Sleep duration is represented by 1, 2 and 3 for sleep perspectives of less than 6 hours, 6-8 hours and more than 8 hours, respectively;
 Nap time is represented by 0 and 1 for taking a nap or not; 
 and Social activity is represented by 0 and 1 for engaging in social activities or not. 
 Cog score is the aggregate score for the combined four different cognition. 
 What we are interested is which factors are related to the depression score. 
 The corresponding model can be represented as follows:
\begin{align*}
y_{ij}=&\beta_0+\beta_1\mbox{Age}_{ij}+\beta_2\mbox{Gender}_{ij}+\beta_3\mbox{Area}_{ij}+\beta_4\mbox{Education}_{ij}
+\beta_5\mbox{Disease}_{ij}\\ +&\beta_6\mbox{Marital status}_{ij}+\beta_7\mbox{Life satisfaction}_{ij}+\beta_8\mbox{Eyesight}_{ij}\\
+&\beta_9\mbox{Hearing}_{ij}+\beta_{10}\mbox{Teeth state}_{ij}+\beta_{11}\mbox{Sleeping time}_{ij}+
\beta_{12}\mbox{Drinking}_{ij}\\+&\beta_{13}\mbox{Nap time}_{ij}+\beta_{14}\mbox{Social activity}_{ij}+
\beta_{15}\mbox{cog score}_{ij}+\epsilon_{ij},
\end{align*}
where $y_{ij}$ is the $j$th depression score of the $i$the subject for $j=1,\cdots,4,$ and $i=1,\cdots, 6796$. 

\begin{figure}[H]
  \centering  
\includegraphics[width=0.8\textwidth]{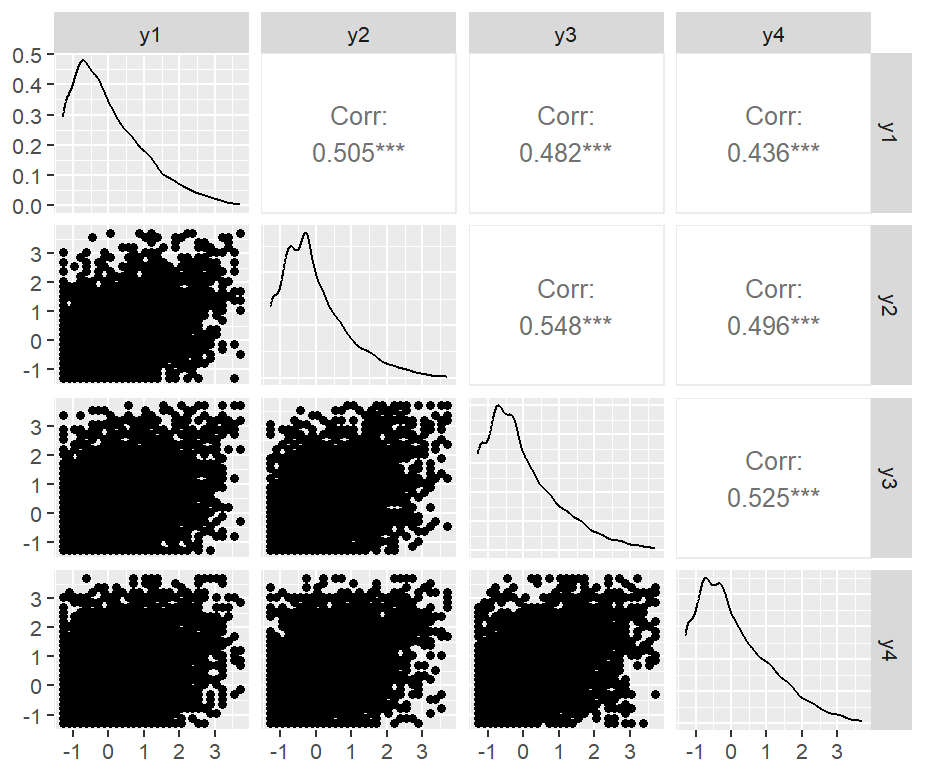}
  \caption{ The scatter plots of the depression, where $y_i$ indicates the $i$th depression score, $i=1,2,3,4$.}
  \label{Fig-real-0}
\end{figure}

\begin{figure}[H]
\centering  
\subfigure[Independent]{
  \includegraphics[width=0.45\textwidth]{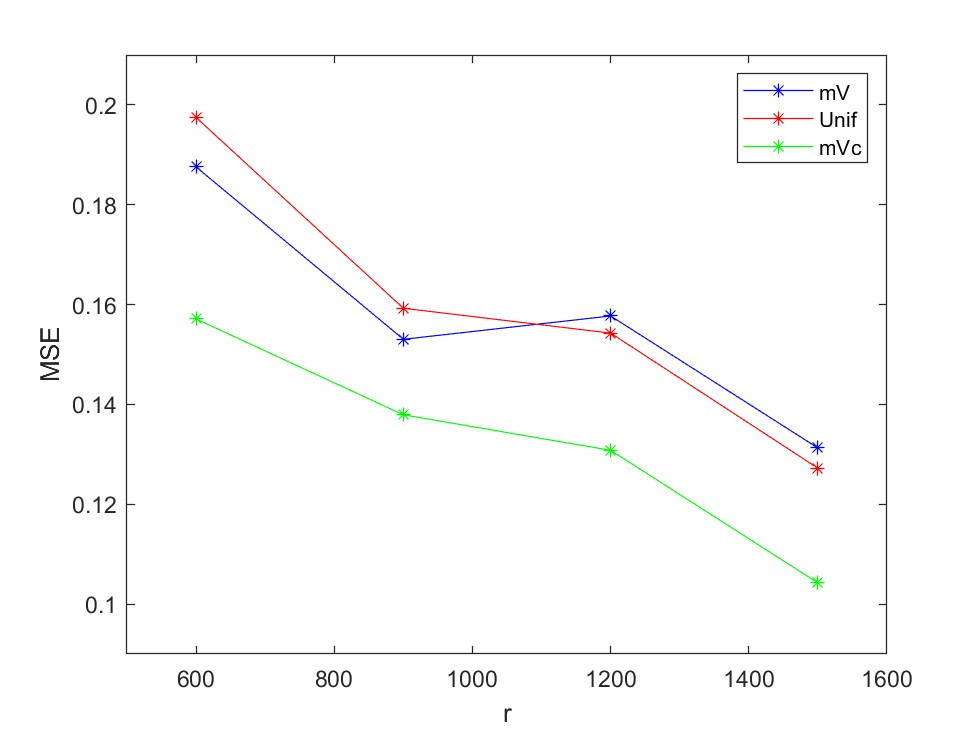}}
    \subfigure[Exchangeable]{
\includegraphics[width=0.45\textwidth]{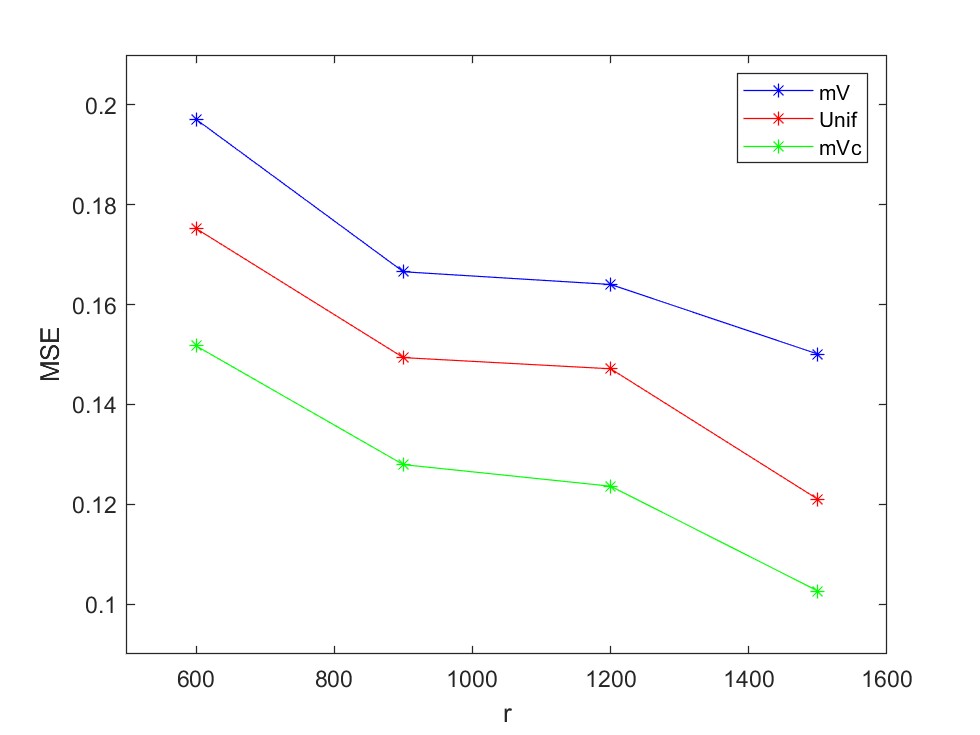}}
    \centering  
\subfigure[AR(1)]{
\includegraphics[width=0.45\textwidth]{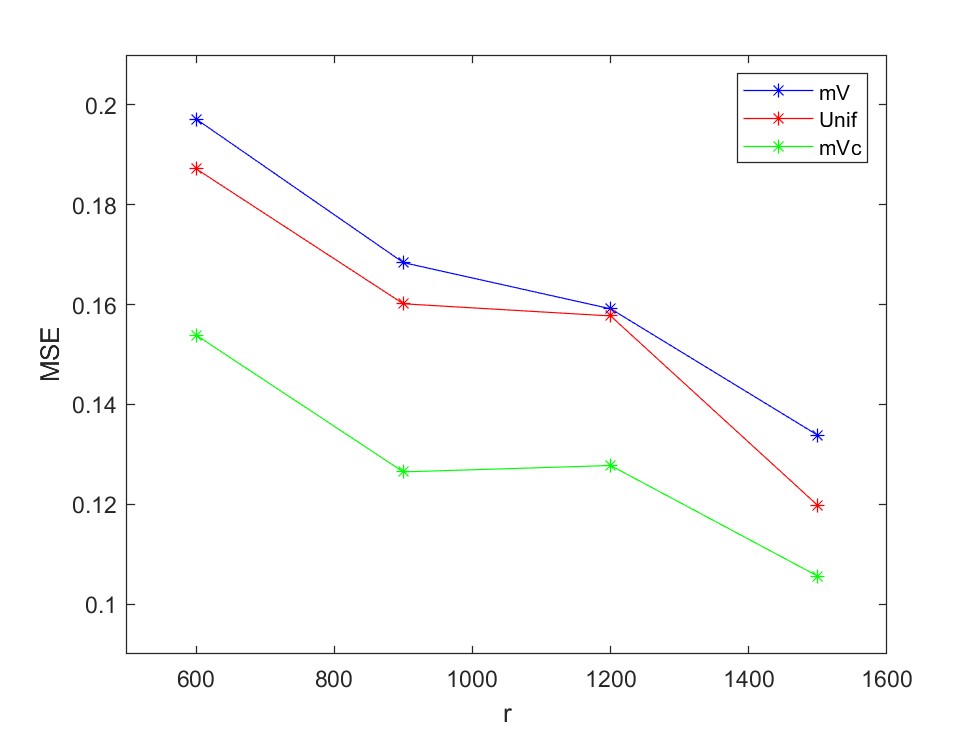}}
 \subfigure[Unstructured]{
\includegraphics[width=0.45\textwidth]{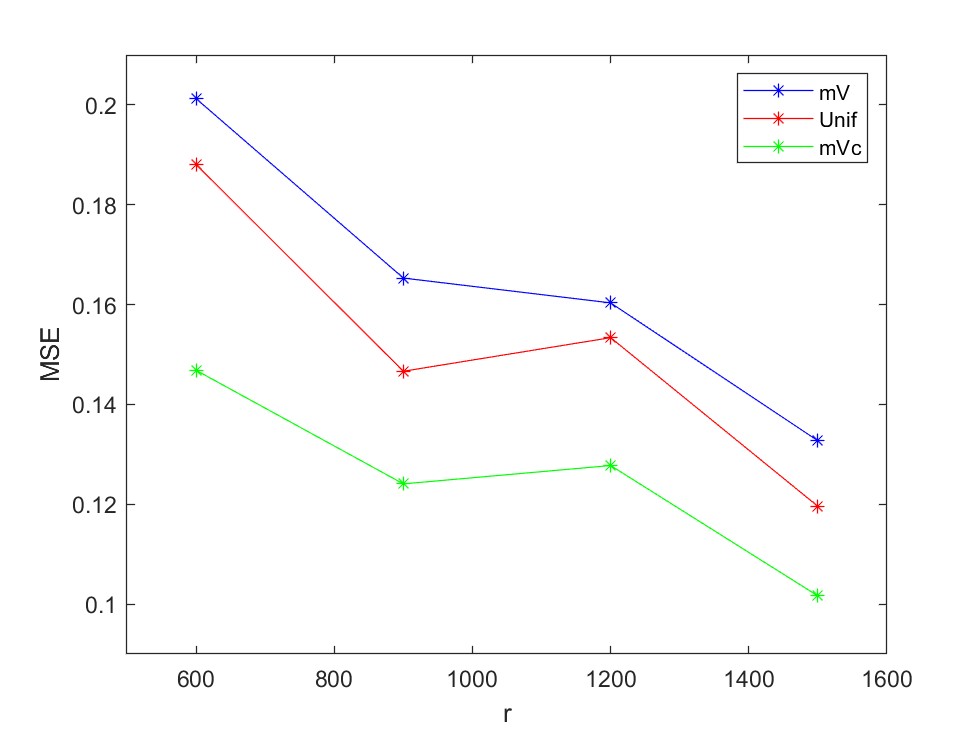}}
\caption{ The MSEs for the real data using different working matrices.}
    \label{Fig-real-1}
  \end{figure}

\begin{figure}[H]
    \centering  
    \subfigure[Independent]{
    \includegraphics[width=0.45\textwidth]{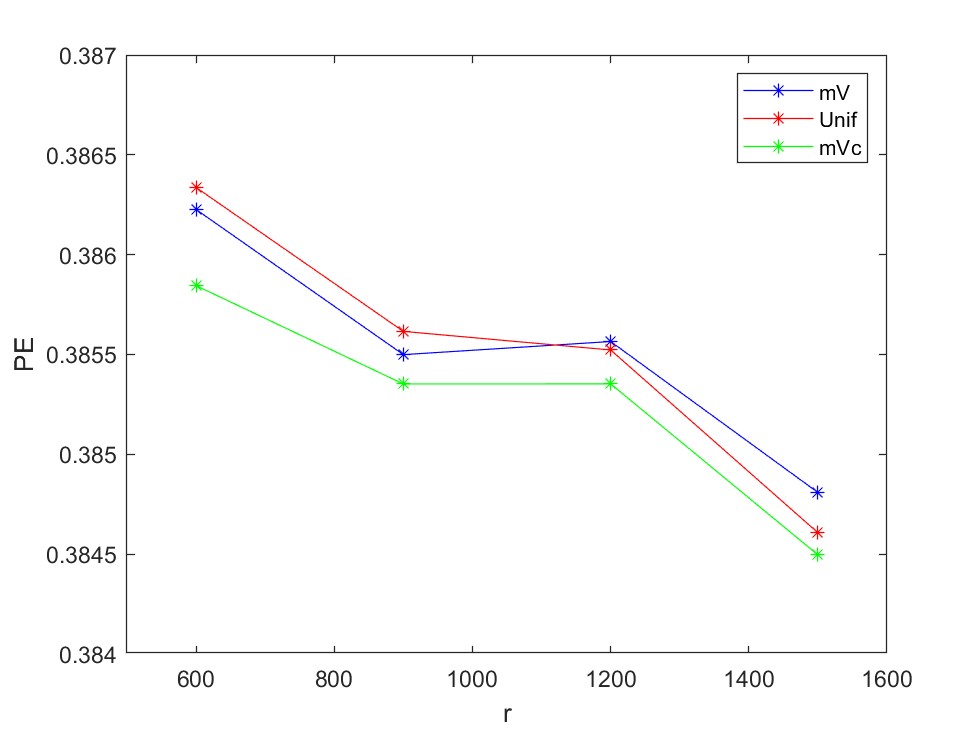}}
    \subfigure[Exchangeable]{
    \includegraphics[width=0.45\textwidth]{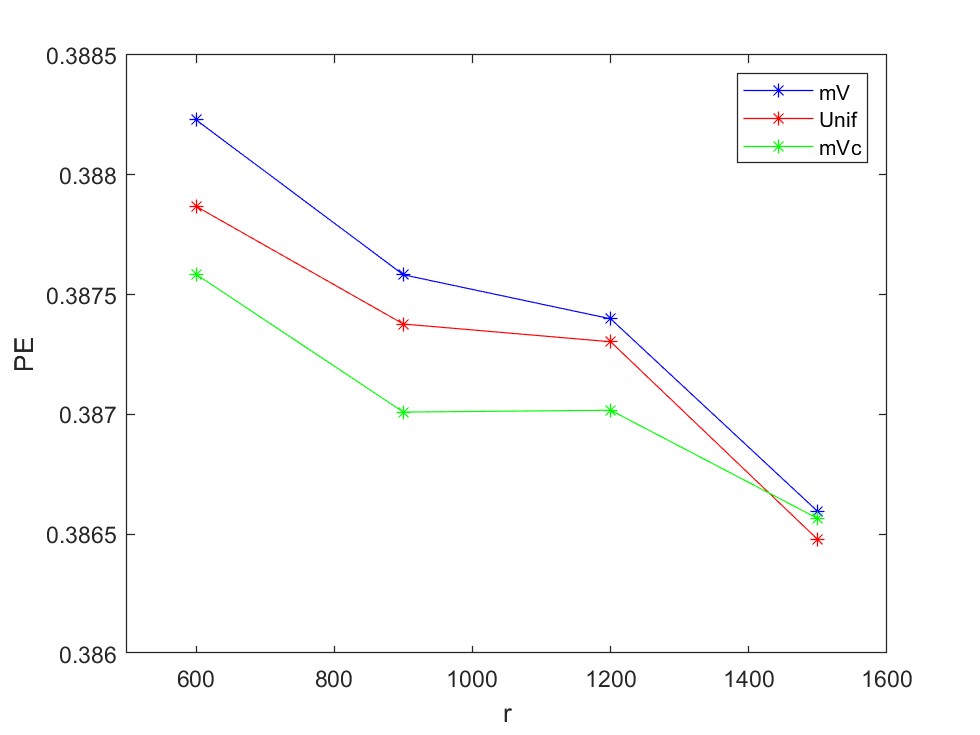}}
    \centering  
    \subfigure[AR(1)]{
    \includegraphics[width=0.45\textwidth]{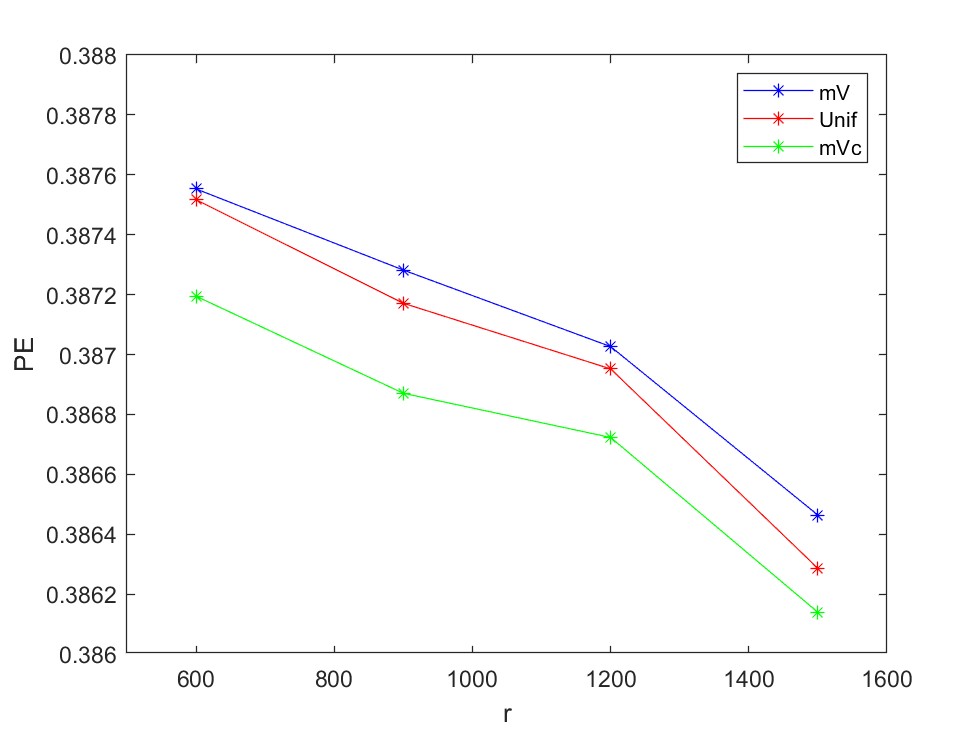}}
     \subfigure[Unstructured]{
    \includegraphics[width=0.45\textwidth]{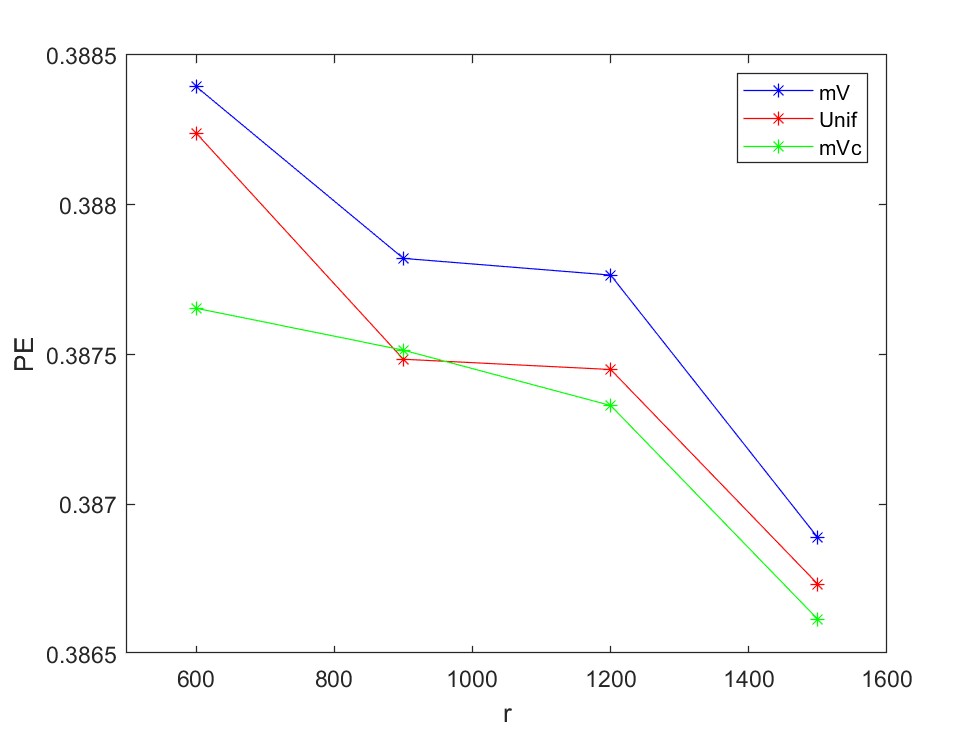}}
    \caption{The prediction errors for three types sampling probabilities under four working correlation matrices.}
    \label{Fig-real-2}
  \end{figure}

Figure \ref{Fig-real-0} is the scatter plot of the depression scores. It is easy to see that there exists moderate correlations. The data is divided into two parts: $5000$ samples are selected as the training set, 
and the left $1796$ samples are as the test set.  Figure \ref{Fig-real-1} shows the MSEs of $\tbeta$ using different methods. 
When calculating MSE, we use $\mbeta$ obtained from the whole data instead of
 the true value. Figure \ref{Fig-real-2} shows the predicted errors (PE) for the test data under different working correlation matrices.
The results show that all three methods fit and predict the data well. 
  The mVc method performs better than the other two methods in different situations. 
  The estimates of the mV method is better than those obtained from the uniform sampling in most cases. When 
  the working  matrix is independent, the results obtained by mV and the uniform 
  sampling are similar. 

\section{Conclusions}
In this paper, we study the optimal subset sampling method for the marginal model with the longitudinal data, and derive the  
optimal sampling probability under the criteria of minimizing the variances of the regression estimators. 
Furthermore, we establish the asymptotic properties of the estimator as the theoretical guarantees.
A various of simulations are carried out for different cases to ensure the feasibility  and the reliability of the proposed method. In this paper, we assume that the dimension of the covariates is fixed. 
However, there are no theoretical guarantees for the case that the dimensionality of the parameters 
diverges with the sample size. Therefore, we will study and explore the problem in the future work.
 In addition, most sampling methods could not be directly extended to the 
 longitudinal data, and how to sampling from the full data within an optimal sampling probability to reduce the computational 
 burden and improve the accuracy of parameter estimators in nonparametric models with massive longitudinal data also need to be further studied.
 
 \section*{Acknowledge}
This research was supported by the National Natural Science Foundation of China (No. 12371295), and Shaanxi Fundamental Science
Research Project for Mathematics and Physics (No. 22JSY002). 
The authors wish to thank the National Development Research Institute of Peking University and  China Social Sciences Survey Center of Peking University for providing CHARLS data.


\begin{thebibliography}{100}

  \bibitem[Ai et al., 2021]{ai2021} Ai, M., Yu, J., Zhang, H. M., and Wang, Y. H. (2021). Optimal subsampling algorithms for big data
  regressions. Statistica Sinica, 31(2), 749--772.
\bibitem[Cram\' er, 1946]{cram46} Cram\' er, H. (1946). Mathematical Methods of Statistics. Princeton University Press.

\bibitem[Crowder, 1986]{crow86} Crowder, M. (1986). On consistency and inconsistency of estimating equations.  Journal of Multivariate Analysis. 65, 245--260.



\bibitem[Diggle et al., 2002] {diggle2002} Diggle, P., Heagerty, P., Liang, K-Y., Zeger S. (2002). Analysis of Longitudinal Data. Oxford University Press.

\bibitem[Drineas et al., 2006]{drin06} Drineas, P., Mahoney, M. W., and Muthukrishnan, S. (2006).  Sampling algorithms for $l_2$ regression and applications. Proceedings of the Seventeenth Annual ACM-SIAM Symposium
  on Discrete Algorithm, 1127--1136.

\bibitem[Drineas et al., 2011]{drin11} Drineas, P., Mahoney, M. W., Muthukrishnan, S., and Sarlós , T. (2011). Faster least squares
  approximation. Numerische Mathematik 117, 219--249.

  \bibitem[Fan et al., 2021]{fan2021}Fan, Y, Liu, Y.K., Zhu, L.X.(2021),
Optimal subsampling for linear quantile
regression models. The Canadian Journal of Statistics  49(4), 1039--1057.
  

\bibitem[Li, 1996]{li96} Li, B. (1996). A minimax approach to consistency and efficiency for estimating equations. The Annals of Statistics 24, 1283--1297.
\bibitem[Liang and Zegar, 1986]{lian86}Liang, K. Y. and Zegar, S. L. (1986). Longitudinal data analysis using generalized
  linear models.  Biometrika 73, 13--22.

\bibitem[Ma and Sun, 2015]{ma15} Ma, P. and Sun, X. (2015). Leveraging for big data regression. Wiley Interdisciplinary
  Reviews: Computational Statistics 7, 70--76.




  \bibitem[Wang,  2019]{wang19a} Wang, H. (2019). Divide-and-conquer information-based optimal subdata selection algorithm. Journal of Statistical Theory and Practice, 19:36.

  \bibitem[Wang and Ma, 2021]{wang:ma:2021}Wang, H., and Ma, Y. (2021). Optimal subsampling for quantile regression in big data. Biometrika 108(1), 99-112.
\bibitem[Wang et al., 2023]{wang2023}Wang, Z., Wang, H. and Ravishanker, N. (2023). Subsampling in longitudinal models. Methodology and Computing in Applied Probability 25:35. 
\bibitem[Wang et al., 2019]{wang18a} Wang, H., Yang, M., and Stufken, J. (2019). Information-based optimal subdata selection for big data linear regression. Journal of the American Statistical Association
  114(525), 393--405.

\bibitem[Wang et al., 2018]{wang18b} Wang, H., Zhu, R., and Ma, P. (2018). Optimal subsampling for large sample logistic
  regression. Journal of the American Statistical Association 113, 829--844.

\bibitem[Yu et al., 2022]{yu2022} Yu, J., Wang,
H. Y.,  Ai, M. Y. , Zhang, H. M. (2022). Optimal distributed subsampling for maximum quasi-likelihood estimators with massive data,  Journal of the American Statistical Association, 117(537), 265--276.




 
  \bibitem[Yuan et al., 1998]{yuan98} Yuan, K. H. and Jennrich, R. I. (1998). Asymptotics of estimating equations under natural
  conditions. Journal of Multivariate Analysis 65, 245--260.

\bibitem[Yuan et al., 2022]{yuan2022} Yuan, X. H., Li, Y., Dong, X. G., Liu, T. Q. (2022). Optimal subsampling for composite quantile regression in big data. 
Statistical Papers 63, 1649--1676.


\bibitem[Zhao et al., 2013] {zhao2013}  Zhao Y. H., Strauss, H., Yang G. H. , Giles H., Hu P. F., Hu Y. S., Lei X. Y., Park A., Smith J. P., Wang Y. F. (2013).
China health and retirement longitudinal study, 2011-2012 National
Baseline Users' Guide, National School of Development, Peking University.

\bibitem[Zuo et al., 2021]{zuo2021} Zuo, L. L.,  Zhang, H. X., Wang, H. Y., Sun, L. Q. (2021). Optimal subsample selection for massive logistic regression
with distributed data.
Computational Statistics 36, 2535--2562.
\end{thebibliography}
\end{document}